\documentclass[11pt,eprint,nofootinbib]{revtex4-2}
\usepackage[utf8]{inputenc}
\usepackage{geometry, bbold, mathrsfs, amssymb, simpler-wick, dsfont, bbm}
\usepackage[colorlinks=true, linkcolor=blue, linktocpage=true, citecolor=blue, urlcolor=blue]{hyperref}
\usepackage[italicdiff]{physics}
\usepackage[normalem]{ulem}
\newcommand{\sh}[0]{\text{Sh}}
\newcommand{\swap}[0]{\text{SWAP}}

\newcommand{\Fint}[1]{\int \mathcal{D}#1 \; }

\newcommand{\reals}[0]{\mathds{R}}
\newcommand{\nats}[0]{\mathds{N}}
\newcommand{\ints}[0]{\mathds{Z}}

\newcommand{\nn}[0]{\nonumber \\}

\newcommand{\qwhere}[0]{\quad \text{where} \quad}

\newcommand{\old}{\color{black}}
\numberwithin{equation}{section}

\begin{document}

\title{Mutual Information of Measurement in Conformal Field Theories}
\author{Israel Klich} 
\author{Ravisankar Rajagopal} 
\author{Diana Vaman} 

\affiliation{Department of Physics,  University of Virginia, \\ Charlottesville, Virginia 22904-4714, USA}

\date{\today}

\begin{abstract}
    The mutual information of measurement of a local operator between two spatial regions characterizes the reduction in the uncertainty of measurement outcomes in one region given the outcome in the other. We study the Rényi-Shannon mutual information of measurement in conformal field theories (CFTs). Explicit calculations for scalar primaries and descendants in large $N$ CFTs are performed in the two extreme limits of the intervals being adjacent or largely separated. The results depend on only the cross ratio and the scaling dimension $\Delta$ of the operator being measured, with striking similarities to the mutual information of entanglement. We find a surprising quadratic dependence as $(\Delta-1/2)^2$ for adjacent intervals. This dependence superficially seems to violate the information processing inequality, which we discuss and resolve. We also offer a holographic interpretation of the mutual information of measurement.
  \end{abstract}

\maketitle
\flushbottom
% \newpage

  \hypersetup{linkcolor=blue}
   \tableofcontents
   
   % \newpage

\maketitle
\section{Introduction}
Entanglement entropy is an important quantity in the study of critical {phenomena}. Although  entanglement entropy is UV divergent, it contains universal information about the underlying system. In $1+1$ dimensional conformal field theories (CFTs), it is known that the von Neumann entanglement entropy derived from the reduced density matrix $ \rho_A$ associated with region $A = [0, \ell]$ takes the form \cite{Holzhey_1994, Calabrese-Cardy}
\begin{equation}
    {S}_A = -\tr( \rho_A \log  \rho_A) = \frac{c}{3}\log(\frac{\ell}{\epsilon}) + \text{constant}
\end{equation}
where $c$ is the central charge. 
In higher even spacetime dimensions, the entanglement entropy likewise contains a universal logarithmic contribution. Unlike in two dimensions, however, this term is subleading to the UV-divergent area-law contribution associated with the boundary of the region.

Entanglement entropy plays an important role in the study of quantum gravity through the AdS/CFT correspondence \cite{Maldacena:1997re, Witten:1998qj}. For a spatial region $A$ in the boundary CFT,  its entanglement entropy is given, at leading order in the semiclassical limit, by the area of a bulk extremal surface anchored on the boundary of $A$
\cite{Ryu:2006bv,Ryu:2006ef}. Changes to the  boundary quantum state, and hence of its entanglement structure, are encoded holographically in corresponding modifications of the bulk geometry, while changes to the boundary region $A$  probe different regions of a given bulk spacetime. 
Extensions of the extremal-surface prescription, including quantum corrections and the notion of an entanglement wedge (the bulk region which can be reconstructed from the boundary data), have strengthened the connections between the holographic bulk spacetime and the entanglement structure of the boundary quantum field  theory \cite{Hubeny:2007xt,Czech:2012bh, JLMS}.

While entanglement entropy characterizes the extent of entanglement of a given quantum state $\rho_A$, there is an analogous quantity, called the Shannon entropy of measurement that characterizes the uncertainty of measurement outcomes for some  local observable. To properly define what it means, consider measuring the field configurations of some local operator  ${\mathcal{O}}(x,t{=}0)$ in a field theory. If $f(x)$ denotes a particular configuration of the operator, by repeated measurements, we can construct a probability distribution $P[f]$ that tells us the probability of measuring the particular outcome $f$. 
For a spatial region $A$, the reduced probability distribution is,
\begin{equation}
    P_A[f_A] = \int \mathcal{D}f_{A^c} \, P[f],
\end{equation}
where $A^c$ is the complement of $A$. We can calculate a similar quantity to that of the entanglement entropy, known as the Shannon measurement entropy, or in general the $n^{\text{th}}$-Rényi Shannon entropy,
\begin{equation}
    \sh_n(\mathcal{O}; A) = \frac{1}{1-n} \log \int \mathcal{D} f_A \; P_A[f_A]^n, \quad n \in \nats
\end{equation}
The usual Shannon entropy is the $n\to1$ limit.  Shannon entropy quantifies the uncertainties in measurement outcomes, given access only to the outcomes in a particular sub-region of space. Shannon and R\'enyi entropies of measurement outcomes have been studied
extensively in quantum many-body systems. Early studies identified
universal subleading contributions to these entropies and related them
to conformal field theory and boundary critical behavior
\cite{Stephan2009,Stephan2011}. Subsequent work established universal
logarithmic scaling of the Shannon and R\'enyi mutual information in
critical quantum chains, while emphasizing the dependence on the
measurement basis and the R\'enyi index
\cite{AlcarazRajabpour2013,Stephan2014,AlcarazRajabpour2014}.
Related definitions based on R\'enyi divergence have also been
investigated \cite{AlcarazRajabpour2015}. More recently, the Shannon
entropy and mutual information of measured bitstrings have been used
to bound and estimate entanglement in quantum simulators
\cite{Meurice2025,Kaufman2025}. These developments motivate a systematic
continuum study of how measurement mutual information depends on the
observable being measured.

For continuous distributions, Shannon entropy is known to be ill-defined (e.g. it is discretization dependent and can be negative). A well-defined quantity can instead be obtained by comparing two probability distributions. This leads to the relative entropy, or Kullback--Leibler divergence \cite{Kullback:1951zyt}. A closely related quantity, the mutual information of measurement outcomes, is also well defined and it measures the correlations between the outcomes of measurements performed in two regions $A$ and $B$ for some observable. Concretely, the R\'enyi mutual information is obtained from the R\'enyi-Shannon entropies as follows:
\begin{equation}
    I_n(\mathcal{O}; A, B) = \sh_n(\mathcal{O}; A) + \sh_n(\mathcal{O}; B) - \sh_n(\mathcal{O}; A \cup B)\label{eq:shannon_mutual_info_def}
\end{equation}
{For 2d CFTs}, with the intervals $A$ and $B$ sharing a common boundary point, the analogously defined mutual information of entanglement is known to be proportional to $\log(1-\eta)$, where $\eta$ is the cross ratio of the intervals\footnote{A similar expression defined in terms of Renyi entanglement entropies of regions $A,B$ and $A\cup B$. This quantity is entanglement mutual information. 
Unlike the mutual information associated with a particular measurement, which captures the classical correlations of  that observable, entanglement mutual information captures the total correlations between the two subsystems, including both classical and quantum correlations.} {}\footnote{\label{fn:crossratio}For four points $a<b<c<d$, the cross ratio is defined as $\eta=\frac{(b-a)(d-c)}{(c-a)(d-b)}.$
In particular, in the case of two  (almost) adjacent intervals of lengths $l_1$ and $l_2$ separated by $\epsilon$ this reduces to $\eta=\frac{l_1 l_2}{(l_1+\epsilon) (l_2+\epsilon)}$ and $1-\eta\simeq \frac{\epsilon(l_1+l_2)}{l_1 l_2}.$} \cite{Calabrese_2009, Calabrese_2011, Furukawa_2009, De_Nobili_2015, flam}. The proportionality constant is  a fraction of the central charge, and it is universal. It is therefore natural to ask whether the mutual information of measurement  has universal features  or not. 

If the mutual information of measurement is to be universal, it can depend on two things:  the central charge, and  some property of the operator being measured. If we measure a conformal primary, then the quantity that characterizes the operator is its conformal dimension. Previous studies on mutual information of measurement point out its possible relation to the central charge \cite{Stephan2014, Koyluoglu2026}. We note that the central charge of the CFT need not necessarily appear in the mutual information. Indeed, if we take a simple Gaussian theory of two {compact} bosons species, we can measure the $U(1)$ current of only one boson. The central charge here is the number of bosons, which can only appear if we simultaneously measure the individual currents for both species. Therefore our expectation is that mutual information needs to depend on the particular characteristics of the chosen observable, namely the scaling dimension. We note that past studies of the mutual information of measurement did not systematically consider this dependence on the scaling dimension. 

In this article, we present a careful study of the mutual information of measurement of single trace scalar primary 
operators and descendants in large $N$ CFT, as well as free fields. In section \ref{sec:sh-formula-derivation}, we derive a simple formula, equation \eqref{eq:shannon-formula}, to compute the second R\'enyi-Shannon entropy of measurement of any local operator. In section \ref{sec:MIinCFT}, we apply this formula to compute the mutual information of measurement for a conformal single trace scalar primary in $1+1$ dimensional large $N$ CFT. 
For two adjacent intervals $A$ and $B$ of lengths $\ell_1$ and $\ell_2$, such that $\ell_{1,2} \gg \epsilon$, where $\epsilon$ is the UV cut-off, we show that the $n$th mutual information of measurement is given by
\begin{equation}
     I_n(\mathcal{O}; A, B) = \frac{1}{2}\qty(\Delta - \frac{1}{2})^2 \log\qty(\frac{\ell_1 \ell_2}{\epsilon (\ell_1 + \ell_2)}), \qquad n \geq 1,
\end{equation}
where $\Delta$ is the scaling dimension of the scalar operator ${ \mathcal{O}}$. Equivalently, $I_n$  can be written in terms of the cross ratio $\eta$
 as
\begin{equation}
    I_n(\mathcal{O}; A, B) = \frac{1}{2}\qty(\Delta - \frac{1}{2})^2 \log\qty(\frac{1}{1-\eta}), \qquad n \geq 1.
\end{equation}
{The independence of $n$ is due to the single trace sector being almost Gaussian in the large $N$ limit.}
 The calculations in section \ref{sec:MIinCFT} are based on the Szeg\H{o} theorem \cite{strongszego} and the Fisher-Hartwig formula \cite{FisherHartwig1968}, which do not generalize easily when the intervals are disjoint. In section \ref{sec:CHMcalculation}, we provide an alternative calculation using CFT methods. This method also opens up a holographic treatment for the mutual information of measurement. Furthermore, the calculation in section \ref{sec:CHMcalculation} can be generalized to disjoint intervals, as shown in section \ref{sec:MIfor-disjoint}. For disjoint intervals, we find that mutual information of measurement is proportional to $\eta^{2\Delta}$. This result shares similarities with the mutual information of entanglement \cite{Cardy_2013}. 
Examples and technical details are relegated to appendices.

\section{Shannon Entropy of Measurement Outcomes of a Local Operator}\label{sec:sh-formula-derivation}
\noindent Consider any local observable ${\mathcal{O}}(x, t)$ in a $d$-dimensional quantum field theory. We construct an ensemble of systems described by a density matrix ${\rho}$, and measure the observable ${\mathcal{O}}$ at some instant of time $t=0$. From many repeated measurements, we can construct the probability of observing a particular eigenvalue distribution $f(x)$. We denote $\hat{\rho}$ the result of averaging over all degrees of freedom not measured by our local observable. Then we can write 
\begin{equation}
    P[f] = \bra{f}\hat{\rho} \ket{f},
\end{equation}
where for simplicity of notation we have suppressed possible degeneracy indices on  $\ket{f}$. 
For clarity, we will first consider measurements in a discrete lattice $\ints^{d-1}$. Let the eigenstates of ${\mathcal{O}}$ be $\ket{f}$,
\begin{equation}
{\mathcal{O}}_{\vec{j}} \ket{f} = f_{\vec{j}} \ket{f},
\end{equation}
where $\vec{j} \in \ints^{d-1}$ labels the lattice site. Here,
\begin{equation}
\ket{f} = \bigotimes_{\vec{j} \in \ints^{d-1}} \ket{f_{\vec{j}}},
\end{equation}
and  $\ket{f_{\vec{j}}}$ are eigenstates of the one-body operator.
The $n^{\text{th}}$- R\'enyi-Shannon entropy of measuring ${\mathcal{O}}$ - denoted by $\sh_n({\mathcal{O}})$ - is then defined as\footnote{We use the shorthand notation $\left[\prod_{\vec{j} \in \ints^{d-1}} \sum_{f_{\vec{j}}}\right] \equiv \int \mathcal{D}f$.}
\begin{equation}
e^{(1-n)\sh_n({\mathcal{O}})} = \left[\prod_{\vec{j} \in \ints^{d-1}} \sum_{f_{\vec{j}}}\right] (P[f])^n \equiv \int \mathcal{D}f \, (P[f])^n.
\end{equation}
We will be interested mostly in $n=2$ entropy:
\begin{align}
e^{-\sh_2({\mathcal{O}})} &= \int \mathcal{D}f \, \bra{f} \hat{\rho} \ket{f}\bra{f} \hat{\rho} \ket{f} \notag\\
&= \int \mathcal{D}f \int \mathcal{D}f' \int \mathcal{D}q \, \exp\left(i \sum_{\vec{j} \in \ints^{d-1}} q_{\vec{j}} (f_{\vec{j}} - f'_{\vec{j}})\right) \bra{f} \hat{\rho} \ket{f} \bra{f'} \hat{\rho} \ket{f'} \notag\\
&= \int \mathcal{D}q \, \left| \operatorname{tr}\left( \hat{\rho} \, e^{i \sum_{\vec{j}} q_{\vec{j}} {\mathcal{O}}_{\vec{j}}} \right) \right|^2.
\end{align}
We can now write down a continuum version,
\begin{equation}
e^{-\sh_2({\mathcal{O}})} = \int \mathcal{D}q \, \left| \operatorname{tr}\left[\hat{\rho} \, \exp(i \int_{\reals^{d-1}} d^{d-1} \vec{x} \, q(\vec{x}) {\mathcal{O}}(\vec{x})) \right] \right|^2.
\end{equation}
The integral over $q$ is now the usual Feynman functional integral. An alternate representation is given by:
\begin{equation}
    e^{-\sh_2({\mathcal{O}})} = \int \mathcal{D}f\, \mathcal{D}f'\, \mathcal{D}q\;
\bra{f}\hat{\rho}\ket{f'}\bra{f'}\hat{\rho}\ket{f}\,
e^{i \sum_{\vec{j}} q_{\vec{j}}(f_{\vec{j}} - f'_{\vec{j}})}
= \int \mathcal{D}q\; \tr\!\left(\hat{\rho}\, e^{i \sum_{\vec{j}} q_{\vec{j}} {\mathcal{O}}_{\vec{j}}}\, \hat{\rho}\, e^{-i \sum_{\vec{j}} q_{\vec{j}} {\mathcal{O}}_{\vec{j}}} \right),
\end{equation}
where we see the explicit appearance of the post-measurement density matrix 
\begin{equation}
\hat{\rho}_{\text{measured}}=   \int \mathcal{D}q\;  e^{i \sum_{\vec{j}} q_{\vec{j}} {\mathcal{O}}_{\vec{j}}}\, \hat{\rho}\, e^{-i \sum_{\vec{j}} q_{\vec{j}} {\mathcal{O}}_{\vec{j}}}.
\end{equation}
which is block diagonal in the eigen-basis of ${\mathcal{O}}$. 
Next imagine that we only measure the operator on a sub-lattice $A \subseteq \ints^{d-1}$. Then the reduced probabilities are
\begin{equation}
P_A[f_A] = \left(\prod_{\vec{j} \in A^c} \sum_{f_{\vec{j}}}\right) P[f] \equiv \int \mathcal{D}f_{A^c} \, P[f].
\end{equation}
Then the second R\'enyi-Shannon entropy of measurement outcomes for operator ${\mathcal{O}}$ in region $A$ is,
\begin{equation}\label{eq:shannon-formula}
e^{-\sh_2({\mathcal{O}}; A)}\! =\! \int \! \mathcal{D}q_A  \left| \operatorname{tr}\left[ \hat{\rho}_A \exp(i \int_A d^{d-1} \vec{x}  q(\vec{x}) {\mathcal{O}}(\vec{x})) \right] \right|^2\!=\! \int \!\mathcal{D}q_A  \left|\expval{\exp(i \int_A d^{d-1} \vec{x}   q(\vec{x}) {\mathcal{O}}(\vec{x}))}\right|^2,
\end{equation}
With $\hat{\rho}_A$ the reduced density matrix corresponding to the sub-lattice $A$.
Some simple calculations to understand and verify this formula are given in Appendix \ref{app:oscillator} and \ref{app:freescalar}. This formula shares certain similarities with the  SWAP by-displacement formula for the second Rényi entanglement entropy \cite{klich2024swaptransposedisplacementsstabilizer}. For contrast and comparison, the SWAP formula is given in Appendix \ref{app:SWAP}. 

\par  As we have already alerted the reader,  Shannon entropy is 
ill-defined for continuous distributions, and it can be negative\footnote{This can be demonstrated with a simple uniform distribution say,
$
p(x) = 
    2, \text{ for } x \in [0,1/2]
$
.Then,
$
\sh_2 = -\log (\int_0^{1/2} d x \; p^2(x)) = -\log(2) <0 
$.}.
On the other hand
\emph{mutual information of measurement}, defined in Eq. \eqref{eq:shannon_mutual_info_def} when $n\rightarrow 1$ is positive. For Gaussian systems the mutual information does not depend on the index, and in particular the second Renyi mutual information is well behaved.
The mutual information of measurement tells us how statistically independent are the measurements of ${\mathcal{O}}$ in the two regions $A$ and $B$. Or, equivalently, how much information is shared between the two regions with respect to $\mathcal O$.

\section{Mutual Information of Measurement in CFT: Adjacent Intervals}\label{sec:MIinCFT}
\noindent Let ${\mathcal{O}}$ be a {scalar} primary field in a CFT, with scaling dimension $\Delta$, and $\rho$ be the ground state. Let us now evaluate the Shannon entropy for measuring this field in a region $A$\footnote{Examples with $\Delta=0$ (free scalar) and $\Delta=1$ (free fermion charge density) are discussed in appendices \ref{app:freescalar} and \ref{app:freefermion}.}. First, we need to evaluate
\begin{equation}\label{eq:expvalueforexpqO}
    \expval{
\exp\qty(i \int d^d X \, q(X){\mathcal{O}}(X))
},
\end{equation}
where $X = (t, \vec{x})$. 
Evaluating this expectation value is, in general, difficult. We will consider CFTs which are either free or have a limit in which higher point correlators can be neglected. Strongly coupled CFTs with $SU(N)$ or $O(N)$ symmetry in the large $N$ limit have this property \cite{Maldacena:1997re, Witten:1998qj}. We have
\begin{equation}\label{eq:CFT-partition-function}
Z[q]=\expval{\exp(i\int_A d^d X \,q(X) \mathcal{O}(X))}=\exp(W[q]),
\end{equation}
where the generating functional of connected correlators is given by
\begin{equation}
W[q]=\sum_n\frac{i^n}{n!}\int dX_1 q(X_1) \dots\int dX_n q(X_n)\langle \mathcal{O}(X_1) \dots \mathcal{O}(X_n)\rangle_c \label{wq}.
\end{equation}
For a single trace operator $\mathcal{O}(x)$ normalized such that its 2-point function is of order $\mathcal O(N^0)$,  each higher-order correlator is suppressed by an additional power of $N$. Therefore, in the large $N$ limit, the leading order contribution to the partition function comes from the 2-point function, making this sector almost Gaussian.
For such a $d$-dimensional CFT, we end up with 
\begin{equation}
    \expval{
\exp\qty(i \int d^d X \, q( X){\mathcal{O}}( X))
} = \exp(- \frac{1}{2}\int d^d X \, d^d  X' \, q(X)
\mathcal{G}_\Delta(X,X')
q(X')),
\end{equation}
where
\begin{equation}\label{eq:two-point-for-O}
\mathcal{G}_\Delta( X, X') = \expval{{\mathcal{O}}( X){\mathcal{O}}( X')}
=
% \frac{(\Delta - d/2)\Gamma(\Delta)}{\pi^{d/2}\Gamma(\Delta - d/2)}
\frac{C_\Delta}{| X -  X'|^{2\Delta}} \, ,
\end{equation}
where we have Wick rotated to Euclidean signature, and $C_\Delta$ a normalization constant. This is a matter of convenience for our measurement problem, since we are interested in simultaneous measurement of the field\footnote{We can simultaneously measure $\mathcal O(\vec x,0)$ in some region only if the commutator $[\hat {\mathcal{O}}( \vec{x},0), \hat {\mathcal{O}(}\vec{y},0)]$ vanishes. However, since $\hat {\mathcal O}$ is a composite operator, there could be non-zero contact terms. For a scalar operator, these terms are forbidden: $\delta(\vec x-\vec y)$ is symmetric while the commutator is antisymmetric under $\vec x, \vec y$ interchange. For $U(1)$ currents, the commutator vanishes as well. We have to be careful in applying the formula \eqref{eq:shannon-formula} for observables like energy-momentum tensor due to such contact terms.}.
Take
\[
q({X}) = \epsilon^{\Delta - d + 1}\, \delta(t)\, q(\vec{x}),
\]
We then obtain,
\begin{align}\label{eq:characteristic functional}
\expval{
\exp\qty(i \int d^d  X \, q( X){\mathcal{O}}( X))
}
&= \exp[
-\frac{1}{2}\int d^{d-1}\!\left(\frac{\vec{x}}{\epsilon}\right)
   d^{d-1}\!\left(\frac{\vec{x}'}{\epsilon}\right)
q(\vec{x})
\left[
C_\Delta
\frac{\epsilon^{2\Delta}}{|\vec{x}-\vec{x}'|^{2\Delta}}
\right]
q(\vec{x}')].
\end{align}
The integral over $q(\vec{x})$ in \eqref{eq:shannon-formula} can now be done, and since $q$ is restricted to the region $A$, we end up with a functional determinant of a translationally invariant kernel {\it restricted to the region $A$}. 
To make the determinant structure explicit, specialize to $d=2$ and
denote the dimensionless equal-time Green's function appearing in
\eqref{eq:characteristic functional} by
\[
\mathcal G_{\Delta,\epsilon}(x-y)
=\epsilon^{2\Delta}\langle \mathcal O(x,0)\mathcal O(y,0)\rangle_c,
\]
with the short-distance regularization understood. If the interval
$A$ contains $M$ lattice points with spacing $\epsilon$, its restriction
to $A$ is the $M\times M$ matrix
\[
(\mathcal G_{\Delta,A})_{ij}
=\mathcal G_{\Delta,\epsilon}\bigl((i-j)\epsilon\bigr),
\qquad i,j=1,\ldots,M.
\]
The Gaussian integration over $q_A$ in \eqref{eq:shannon-formula} therefore gives
\[
e^{-\mathrm{Sh}_2(O;A)}
\propto \bigl[\det\mathcal G_{\Delta,A}\bigr]^{-1/2}.
\]
Thus the entropy is proportional to the logarithm of this restricted
determinant, up to local measure-normalization terms that cancel in
mutual information. Translation invariance makes the matrix elements
depend only on $i-j$, so $\mathcal G_{\Delta,A}$ is a Toeplitz matrix.
Its large-$M$ determinant asymptotics are described by the strong
Szeg\H{o} theorem when the momentum-space Green's function is sufficiently
regular and nonvanishing, and by Fisher--Hartwig asymptotics when it
has algebraic zeros or singularities
\cite{deift2012toeplitzmatricestoeplitzdeterminants}. A physically transparent approach to the asymptotics is derived in \cite{Klich:2026mlr} and more details can be found in appendix \ref{app:Szego}. These methods have been used extensively
in entanglement-entropy calculations for quantum spin chains and
free-fermion systems
\cite{JinKorepin2004,ItsJinKorepin2005}.
Here they allow us to separate the extensive, cutoff-dependent
contribution from the subleading terms relevant to mutual information.
We therefore begin by Fourier transforming the Green's function.
\begin{align}
\mathcal{G}_\Delta(p)
=
\frac{1}{\epsilon}
\int dx\, e^{ipx}
C_\Delta
\frac{\epsilon^{2\Delta}}{ x^{2\Delta}}
=
2 C_\Delta \Gamma(1- 2 \Delta) \sin(\pi \Delta) (\epsilon^2 p^2)^{\Delta-1/2}.
\end{align}
The factor \(1/\epsilon\) keeps the expression dimensionless. We have regularized the Fourier transform by analytic continuation. 
It is useful to think of the system as defined on a lattice with spacing \(\epsilon\). The momenta are then periodic,
\(
p \in [-\pi/\epsilon,\; \pi/\epsilon).
\)
Since \(p\) is periodic, we expect the matrix to respect this periodicity. This is not manifest in the above expression, and we remedy it by replacing $\epsilon p \;\rightarrow\; \sin(\epsilon p).$
Define $\theta = \epsilon p.$
Then the integration measure becomes
$\int_\reals \frac{dp}{2\pi}
\;\rightarrow\;
\frac{1}{\epsilon}
\int_{-\pi}^{\pi} \frac{d\theta}{2\pi}.$
The leading Szegő term is therefore
\begin{align}
\sh_2({\mathcal{O}};A)
=
\frac12
\frac{\ell}{\epsilon}
\int_{-\pi}^{\pi}
\frac{d\theta}{2\pi}
\log
\left(
2 C_\Delta\Gamma(1- 2 \Delta) \sin(\pi \Delta) (\sin^2 \theta)^{ \Delta-1/2}
\right).
\end{align}
The number that multiplies $\ell$ can be computed numerically, and it turns out to be highly cut-off dependent, and therefore not universal. 
\par  
We are interested in the mutual information, and need to compute the sub-leading asymptotic of the determinant for that. 
Introducing an infra-red cut-off $\mu$,
\begin{equation}
\mathcal{G}_\Delta\!\left(\epsilon \abs{p}\right)
\rightarrow
\text{const}\times
\left(\epsilon^2 (p^2+\mu^2)\right)^{\frac{-1+2\Delta}{2}} .
\end{equation}
 $\mu$ will be sent to zero at the end. Now we can use Szeg\H{o} theorem (see appendix \ref{app:Szego}) to get the sub-leading term,

\begin{equation}
   \sh_2({\mathcal{O}};A)
=
\frac{\ell}{\epsilon}\,\#
+
\frac12
\left(\Delta - \frac{1}{2}\right)^2
\log\!\left(\frac{\ell}{\epsilon}\right). \label{sh2}
\end{equation}
The sub-leading term is always positive, and its coefficient is independent of normalization. It is completely determined by the scaling of the two-point function of \({\mathcal{O}}\) given in \eqref{eq:two-point-for-O}.
\par  Let $A$  be an interval of size $\ell_1$, and $B$ of size $\ell_2$ and we take the intervals to be adjacent. Substituting in the definition \eqref{eq:shannon_mutual_info_def} we get the second R\'enyi-Shannon mutual information to be
\begin{equation}\label{eq:I2-result}
    I_2({\mathcal{O}}; A, B) = \frac{1}{2}\qty(\Delta - \frac{1}{2})^2 \log\qty(\frac{\ell_1 \ell_2}{\epsilon (\ell_1 + \ell_2)}).
\end{equation}
The argument of the logarithm is a cross ratio (see footnote \ref{fn:crossratio}). In particular, 
\begin{equation}
    \eta = \frac{\ell_1 \ell_2 }{(\ell_1 + \epsilon)(\ell_2 + \epsilon)}, \quad 0 \leq \eta \leq 1.
\end{equation}
and we have
\begin{equation}\label{eq:I2-O}
    I_2({\mathcal{O}}; A, B) = \frac{1}{2}\qty(\Delta - \frac{1}{2})^2 \log\qty(\frac{1}{1-\eta}).
\end{equation}
The result is similar to the second R\'enyi (entanglement) mutual information between adjacent intervals \cite{Calabrese_2009, Furukawa_2009, De_Nobili_2015, flam}, with the dependence on the central charge being replaced by a dependence on the conformal dimension of the operator which is  measured. 
We emphasize that this result holds for operators whose higher point correlators are small in some limit. Examples of such operators include $U(1)$ currents,
in which case the higher correlators are zero, or single trace operators in large $N$ CFT. 
\par  The result in \eqref{eq:I2-O} is rather counter intuitive, especially for large $\Delta$. For large $\Delta$, we expect mutual information to decrease between two regions, since the correlation between far away points is decreasing faster. However, the information shared between the two intervals is through the boundary between the two intervals, where the correlation increases as $\Delta$ increases. 
Indeed, consider measuring some field $\phi$ of scaling dimension $\Delta$ and measuring another field, $\partial_x \phi $, which has a scaling dimension $\Delta + 1$. 
We might expect the mutual information to be identical: as measuring field configurations of $\phi$ immediately tells us what the field configuration of $\partial_x \phi$ are.
Indeed, the information processing inequality, which we describe below, states that local processing (like taking a derivative) can only decrease mutual information ~\cite{Cover2006}.

To make it clear, assume that space is discrete. If we measure the field at all lattice points $\phi_j$, the derivative is $(\partial_x \phi)_j = \phi_{j+1} - \phi_j$.
Thus, when we partition the space into two intervals with a common boundary, the derivative field at lattice points at the boundary already ``knows" about the field across the boundary. The construction of the derivatives from the field configuration essentially becomes a non-local mapping, and the standard local information processing inequalities do not apply. Essentially the same argument carries through as we increase the order of the derivative. 
Thus, there is a fundamental difference between measuring a field, and then forming a derivative field in each part of the system separately, which can only decrease mutual information, and what we do here, which is to first measure the required degrees of freedom (such as a field derivative), and only then partitioning into $A,B$.
We describe the details in appendix \ref{app-information-processing}. In particular, we explicitly show that for Gaussian random variables the different schemes for measurement and partitioning account exactly for the surprising increase of information when looking at a derivative field.

In contrast, when the intervals do not share a boundary, the intuition that faster correlation decay with increased $\Delta$ leads to a decreasing mutual information proves correct, as we will see in section \ref{sec:MIfor-disjoint}.

\par  An important technical detail is the treatment of the regulator. To arrive at \eqref{eq:I2-result}, we have used analytic continuation to regulate the UV divergence to compute the trace. Of course, this is not the only choice. If we decide to put a UV cut-off instead, changing the Green's function to $\mathcal{G}_\Delta (x) = \frac{C_\Delta}{(\epsilon^2 + x^2)^\Delta}$, then the Fourier transform is regulated to $\mathcal{G}_\Delta(p) \sim \abs{\epsilon p}^{\Delta - 1/2} K_{\Delta-1/2}(\abs{\epsilon p})$. This regulated expression is dominated in the $\epsilon \to 0$ limit by contact terms when $\Delta >1/2$, and loses the CFT scaling behavior. So we must remove the contact terms to retain conformal invariance, and the result \eqref{eq:I2-O} remains intact only then. 
% In short, the result is independent of regularization as long as the regulator does not break conformal invariance. 
To make it more clear, assume that the physical measurement prescription is well defined,
and introduce a lattice spacing $\epsilon$ as a short-distance
regulator for the calculation. In the Gaussian approximation used here,
the measured values of the regulated operator $\mathcal O_\epsilon$ at finitely
many points have a joint Gaussian distribution. Their covariance
matrix is simply the connected, regulated two-point function,
\[
(G_\epsilon)_{ij}
=\langle \mathcal O_\epsilon(x_i)\mathcal O_\epsilon(x_j)\rangle_c,
\qquad x_j=j\epsilon.
\]
Thus the covariance is the two-point kernel appearing in the quadratic
term of the characteristic functional \eqref{eq:characteristic functional}; its inverse appears in the
Gaussian probability weight. The two-point matrix must be positive
semi-definite because the variance of any random variable must be positive. For a translation-invariant
two-point function, let $g_{\Delta,\epsilon}(\theta)$ denote its
momentum-space representation, where $\theta=k\epsilon$ is the
dimensionless momentum, with Fourier convention
\[
(G_\epsilon)_{j k}
=\frac{1}{2\pi}\int_{-\pi}^{\pi}
g_{\Delta,\epsilon}(\theta)e^{i(j-k)\theta}\,d\theta.
\]
Restricting this matrix to a contiguous region $A$ of $M$ sites gives
a Toeplitz matrix, denoted $T_M(g_{\Delta,\epsilon})$, whose entries
are the Fourier coefficients of $g_{\Delta,\epsilon}$. In the
Toeplitz literature, this momentum-space two-point function is called
the ``symbol'' of the matrix. We choose a regularization adapted to
the conformal scaling used in the calculation. For the power-law
cases considered here, a convenient family is
\[
g_{\Delta,\epsilon}(\theta)
=h_\epsilon(\theta)
\left|2\sin\frac{\theta}{2}\right|^{2\Delta-1},
\qquad \Delta>0,
\]
where $h_\epsilon$ is smooth, periodic, and strictly positive.
Since $h_\epsilon(0)$ is finite and nonzero, this prescription retains
the specified conformal small-momentum power while allowing different
completions at $|\theta|\sim1$, corresponding to wavelengths comparable
to the cutoff. Smooth changes of $h_\epsilon$ therefore provide a
natural family of regularizations preserving this scaling.
For two-point kernels of this form, the Fisher-Hartwig theorem gives,
at fixed regulator,
\[
\log\det T_M(g_{\Delta,\epsilon})
=M(\log h_\epsilon)_0
+\left(\Delta-\frac12\right)^2\log M+O(1),
\qquad
(\log h_\epsilon)_0
=\frac{1}{2\pi}\int_{-\pi}^{\pi}
\log h_\epsilon(\theta)\,d\theta.
\]
Consequently, smooth changes of $h_\epsilon$ modify the extensive
and constant terms but leave the logarithmic coefficient unchanged.

\par Remarkably, we observe that  mutual information vanishes at $\Delta = 1/2$. To understand this, note that at this value the Green's function switches from being divergent at small momentum to being divergent at large momentum. At this point, the IR and UV effects conspire to make the mutual information an $\order{1}$ number instead of being dependent on the logarithm of the size of the interval. A mathematical explanation using Szeg\H{o} theorem is in appendix \ref{app:Delta-half}.

\subsection{Replica Interpretation and Generalization to $n$-Rényi-Shannon Entropies}
\noindent We perform the $q$ integral in \eqref{eq:shannon-formula} first to get a replica interpretation.  We get
\begin{equation}
    e^{-\sh_2({\mathcal{O}};A)} = \langle 0, 0| \delta_A ({\mathcal{O}}_1 - {\mathcal{O}}_2)|0,0\rangle,
\end{equation}
where $\delta_A$ denotes a delta-function with support in region $A$ only. The subscripts on ${\mathcal{O}}$ denote the copy of the theory to which they belong. In words, to find the second R\'enyi-Shannon measurement entropy, we take two replicas of the CFT vacuum, and find the partition function with the identification of the required operator along the region $A$.
To contrast this with  Rényi entropy, there one identifies all degrees of freedom along $A$ and computes the partition function. For Rényi-Shannon entropy, we only identify a given operator, the one which is measured, and only at the time when it is measured. Let us write
\begin{equation}
    \delta_A ({\mathcal{O}}_1 - {\mathcal{O}}_2) = \lim_{\xi \to 0} \frac{1}{\sqrt{\det_A (\xi)}} \exp\left[-\frac{1}{2\xi} \int_A dx \; ({\mathcal{O}}_1(x) - {\mathcal{O}}_2(x))^2\right].
\end{equation}
where $\det_A(\xi) = \prod_{x \in A} \xi$. To find the second Shannon entropy, we are therefore computing the partition function on two copies of the CFT with a deformation as shown above. In other words, we compute the partition function with the action,
\begin{equation}
    \mathcal S = \mathcal S_{\text{CFT}_1} + \mathcal S_{\text{CFT}_2} + \frac{1}{2\xi} \int_A dx \; ({\mathcal{O}}_1(x) - {\mathcal{O}}_2(x))^2,
\end{equation}
where $\mathcal S_{\text{CFT}_i}$ are the actions for each of the two copies of the CFTs. It is much simpler to change basis to
\begin{equation}
    {\mathcal{O}}_\pm(x) = \frac{{\mathcal{O}}_1(x) \pm {\mathcal{O}}_2(x)}{\sqrt{2}}
\end{equation}
Perform an analogous change for all the fields of the CFTs. In the large $N$ approximation, we have
\begin{equation}
    \mathcal S_{\text{CFT}_1} + \mathcal S_{\text{CFT}_2} = \mathcal S_{\text{CFT}_+} + \mathcal S_{\text{CFT}_-}, \quad \text{to leading order in } 1/N.
\end{equation}
Therefore, the $+$ theory decouples, 
and we are left with
\begin{equation}
    Z_2 = \frac{1}{\sqrt{\det_A (\xi)}}\left\langle \exp\left(-\frac{1}{\xi} \int_A dx \; {\mathcal{O}}_-^2(x)\right)\right\rangle_- .
\end{equation}
This is known as a double trace deformation and it has been studied, for example,  in \cite{witten2002multitraceoperatorsboundaryconditions, Hartman_2008}. The trick is to introduce an auxiliary field $\sigma$ and change the action of the CFT to, 
\begin{equation}
    \mathcal S \to \mathcal S - \xi \int_A dx \; \left(\sigma -i \frac{{\mathcal{O}_-}}{\xi}\right)^2.
\end{equation}
This is a trivial change, as integration over $\sigma$ is just a Gaussian. Instead of integrating over $\sigma$, we then expand the square, and we see that it cancels the double trace deformation. Therefore,
\begin{equation}
    Z_2 =  \int \mathcal{D}\sigma \left\langle \exp\left[-\xi \int_A dx \; \sigma^2(x) + \int_A dx \;2 i \sigma(x)\,{\mathcal{O}_-}(x)\right]\right\rangle .
\end{equation}
the $\det_A(\xi)$ in the denominator gets canceled from the same factor needed for normalization of the $\sigma$ path integral. We will now assume the large $N$ limit, and compute the expectation value to the lowest order. So,
\begin{align}
\begin{split}
    Z_2 &= \int \mathcal{D}\sigma \; \exp\left(-\xi \int_A dx \; \sigma^2(x)\right)
    %\\ &\times 
    \exp\left[- \int_A dx\, dy \; \sigma(x)\langle{\mathcal{O}_-}(x){\mathcal{O}_-}(y)\rangle\sigma(y)\right]
\end{split}\nonumber\\
    &= \frac{1}{\sqrt{\det_A[{2(\xi + \hat{\mathcal{G}}_\Delta})]}}.
\end{align}
Therefore, the second Rényi-Shannon entropy  is
\begin{equation}
    \sh_2({\mathcal{O}}; A) = -\log Z_2 \underset{\xi \to 0}{ =} \frac{1}{2}\operatorname{tr}_A \log\!\left({2\mathcal{G}}_\Delta\right) + O(\xi)
    \label{sh2szegoxi}.
\end{equation}
This is identical to the previous result.
\par 
The advantage of this method is that we can generalize it to higher Rényi-Shannon entropies, for $n > 1$. For the $n$-th entropy, we make $n$ copies of the vacuum, and identify the operator cyclically. That is,
\begin{equation}\label{eq:cyclic-identification}
    e^{-(n-1)\sh_n({\mathcal{O}};A)} = \langle 0|^{\otimes n}\left[\delta_A({\mathcal{O}}_1 - {\mathcal{O}}_2)\,\delta_A({\mathcal{O}}_2 - {\mathcal{O}}_3)\cdots\delta_A({\mathcal{O}}_{n-1} - {\mathcal{O}}_n)\right]|0\rangle^{\otimes n}.
\end{equation}
We can perform an orthogonal transformation that decouples all the replicas. Define a new basis,
\begin{equation}
    \widetilde{\mathcal{O}}_n = \frac{1}{\sqrt{n}} \sum_{i=1}^n \mathcal{O}_i, \;\; \text{and}\;\; \widetilde{\mathcal{O}}_j = \frac{1}{\sqrt{j(j+1)}}\qty(\sum_{i=1}^j \mathcal{O}_i - j \mathcal{O}_{j+1}), \;\; j = 1, 2, \dots, n-1
\end{equation}
Imposing $\widetilde{\mathcal{O}}_j$ fields vanish on $A$ for $j=1,2,\dots, n-1$ is equivalent to the cyclic identification imposed in \eqref{eq:cyclic-identification}. 
% Writing all delta functions as before, and changing to a new basis following \cite{Stephan2014},
% \begin{equation}
%     {\widetilde{\mathcal{O}}}_1 = \frac{1}{\sqrt{n}}\sum_i {\mathcal{O}}_i, \qquad {\widetilde{\mathcal{O}}}_i = \frac{1}{\sqrt{2}}({\mathcal{O}}_i - {\mathcal{O}}_{i-1}), \quad i = 2, 3, \dots, n.
% \end{equation}
The field $ {\widetilde{\mathcal{O}}}_n$ has free boundary conditions on $A$. In the large $N$ limit, the two point functions dominate, and therefore the theory approximately decouples for each $ {\widetilde{\mathcal{O}}}$ field. The change to new variables introduces a Jacobian factor for the delta functions, which can be found from the inverse transformations, 
\begin{equation}
    \mathcal{O}_1 - \mathcal{O}_2 = \sqrt{2} \widetilde{\mathcal{O}}_1, \;\; \text{and} \;\; \mathcal{O}_j - \mathcal{O}_{j+1} = \sqrt{\frac{j+1}{j}} \widetilde{\mathcal{O}}_{j} - \sqrt{\frac{j-1}{j}} \widetilde{\mathcal{O}}_{j-1}\;\; \text{for } j > 1.
\end{equation}
The delta functions therefore pick up the factor,
\begin{equation}
    \prod_{j=1}^{n-1}\sqrt{\frac{j}{j+1}} = \frac{1}{\sqrt{n}}
\end{equation}
After changing the variables to the $\widetilde{\mathcal{O}}$ fields,
we can apply the same double trace deformation procedure for all the $n - 1$ copies, and obtain, 
\begin{equation}
    Z_n = \frac{1}{\sqrt{n}}\left[\frac{1}{\sqrt{ \det_A[2(\xi + {\mathcal{G}}_\Delta})]}\right]^{n-1}.
\end{equation}
Taking a log and taking $\xi \to 0$ limit, we get,
\begin{equation}
    \sh_n(\mathcal{O}; A) = \frac{1}{2} \log \det_A (2\mathcal{G}_\Delta) + \frac{\ell}{2(n-1)} \log n
\end{equation}
The extensive term that depends on $n$ will cancel when computing the mutual information.
Therefore, the $n$th mutual information of measurement $I_n$ is independent of $n$ for Gaussian/large $N$ CFTs. That is,
\begin{equation}
    I_n({\mathcal{O}}; A, B) = \frac{1}{2}\qty(\Delta - \frac{1}{2})^2 \log(\frac{1}{1-\eta}), \quad \forall n \geq 1.
\end{equation}
This agrees with results of \cite{Stephan2014} for free fields
when specializing to conformal dimension $\Delta = 0$. 

Although the result above is independent of the R\'enyi index in the
Gaussian approximation, the second R\'enyi-Shannon entropy has a useful special
property once non-Gaussian corrections are included.  To see this, write
the characteristic functional as in \eqref{eq:CFT-partition-function} 
and write $W[q]$ 
defined in \eqref{wq} succinctly as
\begin{equation}
    W[q]
    =
    \sum_{m=1}^{\infty}
    \frac{i^m}{m!}\,
    C_m[q,\ldots,q].
\end{equation}
For a Hermitian observable, $Z[-q]=Z[q]^*$.  The second R\'enyi-Shannon 
entropy therefore involves
\begin{equation}
    e^{-\sh_2}
    =
    \int Dq\, Z[q]Z[-q]
    =
    \int Dq\, e^{W[q]+W[-q]} .
\end{equation}
All terms in $W[q]$ containing an odd connected correlator are odd
under $q\rightarrow -q$ and hence cancel identically in the combination
$W[q]+W[-q]$.  Thus $\sh_2$ is completely insensitive to
$C_3,C_5,\ldots$.

This observation is particularly useful in a large $N$ CFT.  If the
connected correlators scale schematically as
\begin{equation}
    C_2\sim O(1),\qquad
    C_3\sim O(N^{-1}),\qquad
    C_4\sim O(N^{-2}),\ldots,
\end{equation}
the probability distribution itself receives a non-Gaussian correction
already at order $1/N$, whereas this correction is absent from $\sh_2$.
The leading correction to the Gaussian result for $\sh_2$ therefore
begins at order $1/N^2$, through the connected four-point function.

This provides an additional reason for concentrating on the second
R\'enyi entropy.  At Gaussian order it represents the whole family of
R\'enyi-Shannon entropies, since the result is independent of $n$,
while beyond Gaussian order it enjoys an exact cancellation of all odd
connected correlators.  We stress, however, that this does not imply
that $\sh_2$ is universally closer to the Gaussian result than the
Shannon entropy $\sh_1$; rather, $n=2$ is specially protected against
odd-cumulant corrections.

\section{Mutual Information of Measurement from Holography: Adjacent Intervals}\label{sec:CHMcalculation}

\noindent Consider a $d=2$ CFT and ${\mathcal{O}}(x,t)$ a scalar operator of conformal dimension $\Delta$. Its two-point correlation function, fixed by conformal invariance  is given in \eqref{eq:two-point-for-O}.
In particular, the equal-time two-point correlation function is
\begin{equation}
\langle {\mathcal{O}}(x,0) {\mathcal{O}}(y, 0) \rangle = \frac{C_\Delta}{|x-y|^{2\Delta}} \equiv \mathcal G_\Delta(x-y)\label{equal time}
\end{equation}
 Here we offer an alternative derivation of the spatially restricted functional determinant of the Green's function $\mathcal G_\Delta(x-y)$ using CFT and holographic methods which are applicable beyond the Gaussian fixed point.  We will see that these methods extend to disjoint intervals when there are no analytic results that can be derived using Toeplitz matrix results.

\par  We recall first the Casini-Huerta-Myers conformal map \cite{Casini:2011kv},  
\begin{equation}
t=R\frac{\sinh(\tau/R)}{\cosh u + \cosh (\tau/R)},\qquad r=R\frac{\sinh u}{\cosh u+\cosh (\tau/R)} \label{chy map}
\end{equation}
which maps a spherical spatial region in $d$ dimensional flat spacetime $ds^2=-dt^2+dr^2+r^2 \,d\Omega_{d-2}^2$ to hyperbolic space $\mathds{H}^{d-1}$ (up to a conformal factor)
\begin{equation}
ds^2= (\cosh(u)+\cosh(\tau/R))^{-2}(-d\tau^2+R^2(du^2+\sinh^2 u \,d\Omega_{d-2}^2)).
\end{equation}
We specialize to the $d=2$ case and take the region $A$ to be the interval $(-R,R)$.  Define $r=x$ and allow $x$ to be a spatial coordinate on the real line
\begin{equation}
ds^2=-dt^2+dx^2= (\cosh(u)+ \cosh(\tau/R ))^{-2} (-d \tau^2 + R^2 du^2).
\end{equation}
For $t=\tau=0$, the $x\to u$ map  sends the $(-R,R)$ interval to the real line,
\begin{equation}
x=R\tanh(u/2) \quad \iff \quad u=\log\frac{R+x}{R-x}.
\end{equation}
Under this map, the scalar primary transforms as 
\begin{equation}
\mathcal{O}(u,\tau)=R^\Delta(\cosh(u)+ \cosh( \tau/R ))^{-\Delta} \mathcal{O}(x(\tau,u),t(\tau,u)).
\end{equation}
In particular,
\begin{equation}
\mathcal{O}(u,0)=\bigg|\frac{dx}{du}\bigg|^\Delta \mathcal{O}(x(u),0)\label{o change}
\end{equation}
and so, the vacuum CFT 2-point  function \eqref{equal time} is mapped to
\begin{equation}\label{eq:thermal-2pt}
\langle \mathcal{O}(u,0) \mathcal{O}(v,0) \rangle =\frac{C_\Delta}{2 ^{2\Delta}|\sinh(\frac{v-u}2)|^{2\Delta}}.
\end{equation}
This is a thermal equal-time correlation function. The advantage of this expression lies in the fact that the more difficult problem of computing the restricted determinant
$\det(\mathcal G_\Delta(x;y))$ with $x,y \in (-R,R)$ has been mapped to an unrestricted thermal determinant: $\det(\abs{\sinh(\frac{v-u}{2})}^{-2\Delta})$.
The latter can be evaluated by  diagonalizing it in Fourier space: 
\begin{multline}
\log \det(\abs{\sinh(\frac{v-u}{2})
}^{-2\Delta})= \tr \log \abs{\sinh(\frac{v-u}{2})}^{-2\Delta}\\ =\text{Vol}( \mathds{H}^1)\int \frac{dk}{2\pi} \, \log \bigg(2^{2\Delta+1/2}\pi ^{-3/2} \sin (\pi  \Delta) \Gamma (1-2 \Delta) \cosh \left({\pi  k}\right) \Gamma \left(\Delta-{i k}\right) \Gamma \left(\Delta+{i k}\right)\bigg),
\end{multline}
where $\text{Vol}( \mathds{H}^1)$ is the range of the spatial coordinates $u,v$.
However, this expression is divergent.  We regularize by subtracting the zero-temperature determinant. Recall from \cite{Casini:2011kv} that $T=1/(2\pi R)$, and so with $R \to \infty$ we are subtracting  the log of the unrestricted determinant, with the region $A$ replaced by the whole spatial slice:
\begin{align}
\log \det(\left|\frac{u-v}2\right|^{-2\Delta})=\text{Vol}( \mathds{H}^1)\int \frac{dk}{2\pi} \log \sqrt{\frac{2}{\pi }}2^{2\Delta} \sin (\pi  \Delta) \Gamma (1-2 \Delta) | k| ^{2 \Delta-1}.
\end{align}
The regularized determinant is: 
\begin{align}
\text{Vol}( \mathds{H}^1)\int \frac{dk}{2\pi} \log\bigg(\frac{\cosh\left({\pi k}\right)\Gamma \left(\Delta-{i k}\right) \Gamma \left(\Delta+{i k}\right)}{\pi |k|^{2\Delta-1}}\bigg)\label{reg}.
\end{align}
This is a convergent expression that depends only on $\Delta$. Using that $\Gamma(1/2 - i w)\Gamma(1/2+i w) = \pi/\cosh(\pi w)$ we further notice that for $\Delta=1/2$ the argument of the logarithm is 1.  The integral in \eqref{reg} is a function of $\Delta$ which vanishes for $\Delta=1/2$.  It turns out that the derivative with respect to $\Delta$ of the integral in \eqref{reg} can be computed easily, and it is equal to  $(2\Delta-1)/2$.  Integrating back with respect to $\Delta$ and using that it must vanish for $\Delta=1/2$ gives the regularized determinant
\begin{align}
\log \det(\abs{\sinh(\frac{v-u}2)}^{-2\Delta})-\log \det (\left|\frac{v-u}2\right|^{-2\Delta})=\frac12 \left(\Delta-\frac 12\right)^2 \text{Vol}( \mathds{H}^1).
\end{align}
At this point we recall that the UV regulator in the original $x$ coordinates is mapped into an IR regulator in the $u$ coordinates \cite{Casini:2011kv}.  Concretely, with $-R+\varepsilon\leq x\leq R-\varepsilon$ we have $|u|<\log(2R/\varepsilon)$ which gives a regulated $\text{Vol}( \mathds{H}^1)=2\log(2R/\varepsilon)$.
Putting all of this together,
\begin{equation}
    \sh_2(\mathcal{O};A) - \sh_2(\mathcal{O}, \reals) = \frac{1}{2} \log \det(\frac{\abs{\sinh(\frac{v-u}2)}^{-2\Delta}}{\left|\frac{u-v}2\right|^{-2\Delta}}) = \frac 12 \qty(\Delta-\frac 12)^2 \log(\frac{2 R}{\epsilon})%+\text{local\;terms}
\end{equation}
and we recover the result derived earlier in equation \eqref{sh2}.

In large $N$ CFT,  we can go beyond the two-point function contribution to \eqref{eq:expvalueforexpqO}. From \eqref{eq:CFT-partition-function} and \eqref{wq}, we see that
the first order correction arises from 3-point correlations:
\begin{align}
Z[q]&=\exp\qty(\frac{-1}2 \int dx dy  q(x) \langle \mathcal{O}(x) \mathcal{O}(y)\rangle q(y)) \nn 
& \qquad  \times \bigg[1+\frac {i^3}{3!}\int dx_1 dx_2 dx_3\;  q(x_1) q(x_2) q(x_3) \langle \mathcal{O}(x_1) \mathcal{O}(x_2) \mathcal{O}(x_3)\rangle_c +\dots \bigg]\label{expz}
\end{align}
However, for the computation of the second  Rényi-Shannon entropy we need $|Z[q]|^2=Z[q]Z[-q]$. 
We argued earlier that due to the reality of the observable $\mathcal O$ we have 
\begin{equation}
    |Z[q]|^2=\exp(W[q]+W[-q]).
\end{equation}
This makes it clear that the contribution of all odd connected correlators to the second R\'enyi-Shannon entropy vanishes. Therefore, for large $N$ CFT, the leading non-Gaussian correction to $\sh_2$
arises from the connected four-point function, which is of
order $1/N^2$.

Using the holographic duality between  large $N$ $d=2$ CFTs and supergravity on $\text{AdS}_{3}$, it is standard to compute such correlations via Witten diagrams. Indeed, in the early days of AdS/CFT \cite{Maldacena:1997re} the 3-point and 4-point scalar primary correlation functions were among the first to be computed  \cite{Witten:1998qj,Gubser:1998bc,Freedman:1998tz}.  Here we have to contend with the fact that the sources $q(x)$ need to be integrated over the restricted region $x\in A$. We can proceed  by mapping the interval $A$ to the line. When we extend this map into the bulk AdS as in \eqref{chy map},  the bulk becomes a topological black hole \cite{Casini:2011kv}.  The correlation functions of the conformally transformed operators can be evaluated using Witten diagrams in the black hole background \cite{witten1998antidesitterspacethermal, Barnes:2010jp}. We leave this extension for future work.
Instead,  we focus our attention on the case of disjoint intervals,  and continue to work in the Gaussian approximation in the next section.

\section{ Mutual Information of Measurement: Disjoint Intervals}\label{sec:MIfor-disjoint}
\noindent For disjoint intervals $(-b,-a)\cup (a,b)$ we will use the map \cite{Arias_2018},
\begin{equation}
\exp(u)=-\frac{(x+b)(x-a)}{(x+a)(x-b)}.
\end{equation}
Note that this is a one to two map: for each $u\in \reals $ there are two $x$ images, one in each interval.  We denote them by $x_\pm$, where
\begin{align}
x_\pm=\frac 12\bigg( (b-a) \tanh(u/2)\pm\sqrt{(b-a)^2\tanh^2(u/2) +4 ab}\bigg).
\end{align}
In particular, these images are related by an inversion:
\begin{equation}
x_+ x_- =-a b .
\end{equation}
We further  define
\begin{equation}
\tan(\theta_\pm(u))=\frac{x_\pm(u)}{\sqrt{ab}},\qquad \tan(\theta_+)=-\frac{1}{\tan(\theta_-)}, \qquad \theta_+(u)=\theta_-(u)+\frac \pi 2\label{theta}.
\end{equation}
The cross-ratio for this symmetric choice of disjoint intervals is 
\begin{equation}
\eta=\frac{(b-a)^2}{(b+a)^2}.
\end{equation}
As the intervals move farther apart (in the limit $b-a$ kept fixed as $b\to \Lambda+ b$ and $a\to \Lambda +a$ with $\Lambda\gg1$),  the cross ratio approaches zero and when the intervals move closed together ($a\to a/\Lambda $ with $\Lambda\gg 1$), the cross ratio approaches 1. In terms of $\eta$, we can rewrite
\begin{equation}
\tan(\theta_\pm(u))=\frac{1}{\sqrt{1-\eta}}\bigg(\sqrt\eta \tanh(u/2)\pm\sqrt{1-\eta+\eta\tanh^2(u/2)}\bigg)\label{theta eta}.
\end{equation}
For two disjoint intervals we need to compute a determinant corresponding to the matrix of equal-time correlations 
\begin{align}
\begin{pmatrix}
\langle \mathcal{O}(x_+)\, \mathcal{O}(y_+) \rangle & \langle \mathcal{O}(x_+) \,\mathcal{O}(y_-)\rangle \\
\langle \mathcal{O}(x_-) \,\mathcal{O}(y_+)\rangle & \langle \mathcal{O}(x_-) \,\mathcal{O}(y_-)\rangle 
\end{pmatrix}.
\end{align}
After the map the correlation matrix is no longer Toeplitz. Instead it becomes
\begin{equation}
\begin{pmatrix}
 |\frac{dx_+}{du}|^{\Delta} |\frac{dy_+}{dv}|^\Delta \langle \mathcal{O}(x_+(u))\, \mathcal{O}(y_+(v)) \rangle & |\frac{dx_+}{du}|^{\Delta} |\frac{dy_-}{dv}|^\Delta \langle \mathcal{O}(x_+(u)) \,\mathcal{O}(y_-(v))\rangle \\
|\frac{dx_-}{du}|^{\Delta} |\frac{dy_+}{dv}|^\Delta \langle \mathcal{O}(x_-(u)) \,\mathcal{O}(y_+(v))\rangle & |\frac{dx_-}{du}|^{\Delta} |\frac{dy_-}{dv}|^\Delta \langle \mathcal{O}(x_-(u)) \,\mathcal{O}(y_-(v))\rangle 
\end{pmatrix},
\end{equation}
which evaluates to 
\begin{align}
\frac{C_\Delta}{2^{2\Delta} |\sinh(\frac{v-u}2)|^{2\Delta}}
 \begin{pmatrix}
|\cos(\theta_+(u)-\theta_+(v))|^{2\Delta}&|\cos(\theta_+ (u)-\theta_-(v))|^{2\Delta}\\
|\cos(\theta_- (u)-\theta_+(v))|^{2\Delta}&|\cos(\theta_-(u)-\theta_-(v))|^{2\Delta}
\end{pmatrix},
\end{align}
or, in terms of $\theta_+(u), \theta_+(v)$ only,
\begin{align}
\frac{C_\Delta}{2^{2\Delta} |\sinh(\frac{v-u}2)|^{2\Delta}}
 \begin{pmatrix}
|\cos(\theta_+(u)-\theta_+(v))|^{2\Delta}&|\sin(\theta_+ (u)-\theta_+(v))|^{2\Delta}\\
|\sin(\theta_+ (u)-\theta_+(v))|^{2\Delta}&|\cos(\theta_+(u)-\theta_+(v))|^{2\Delta}
\end{pmatrix}\label{pm}.
\end{align}
Substituting \eqref{theta eta} and expanding in small $\eta$ gives
\begin{align}
\frac{C_\Delta}{2^{2\Delta} |\sinh(\frac{v-u}2)|^{2\Delta}}
 \begin{pmatrix}
1-\frac {\Delta\eta}4|\frac{\sinh(v/2-u/2)}{\cosh(u/2)\cosh(v/2)}|^{2}+\mathcal O(\Delta\eta^2)
&\eta^\Delta 2^{-2\Delta} |\frac{\sinh(v/2-u/2)}{\cosh(u/2)\cosh(v/2)}|^{2\Delta}+\mathcal O(\eta^{\Delta+1})\\
\eta^\Delta 2^{-2\Delta} |\frac{\sinh(v/2-u/2)}{\cosh(u/2)\cosh(v/2)}|^{2\Delta}+\mathcal O(\eta^{\Delta+1})&
1-\frac {\Delta\eta}4|\frac{\sinh(v/2-u/2)}{\cosh(u/2)\cosh(v/2)}|^{2}+\mathcal O(\Delta\eta^2)
\end{pmatrix}\label{pm1}
\end{align}
This makes it clear that the off-diagonal entries, responsible for the correlations between the two intervals are of order $\eta^{\Delta}$  in the limit of small $\eta$ (that is,  for far separated disjoint intervals).  We therefore expect that mutual information will be of order $\eta^{2\Delta}$. 

\par  As an example, we will work out the case $\Delta=1$.  
First we diagonalize the matrix the of \eqref{pm} in $(+,-)$ space,
\begin{equation}
    \frac{C_1}{4 \abs{\sinh(\frac{u-v}{2})}^2}\mqty(1 & 0 \\ 0 & \cos[2(\theta_+(u) - \theta_+(v))]) \equiv \mqty(\bra{u} G_T \ket{v} & 0 \\ 0 & \bra{v} M \ket{u}).
\end{equation}
Here, $G_T$ is the thermal Green's function in \eqref{eq:thermal-2pt}. By insertion of identity in the form of $G_T G_T^{-1}$, we have,
\begin{equation}
    \det \mqty(G_T & 0 \\ 0 & M) = \det\mqty(G_T & 0 \\ 0 & G_T) \det\mqty(1 & 0 \\ 0 & G^{-1}_T M).
\end{equation}
The first determinant equal to $(\det G_T)^2$ is known, based on a computation  we did in the previous section. For the second determinant, we begin by Taylor expanding in small $\eta$, to get
\begin{align}
    \bra{u} M \ket{v} &= \frac{C_1}{4 \abs{\sinh(\frac{u-v}{2})}^2} \qty(1 - \frac{\eta}{2}\qty(\frac{\sinh(\frac{u-v}{2})}{\cosh(\frac{u}{2})\cosh(\frac{v}{2})})^2 + \order{\eta^2})\nn
    &\equiv \bra{u} G_T \ket{v} + \eta \bra{u} M_1 \ket{v} + \frac{\eta^2}{2} \bra{u} M_2\ket{v} + \order{\eta^3},
\end{align}
where we have introduced the shorthand notation 
\begin{equation}
    \bra{u} M_1 \ket{v} = -\frac{C_1}{4 \abs{\sinh(\frac{u-v}{2})}^2}\times \frac{1}{2}\qty(\frac{\sinh(\frac{u-v}{2})}{\cosh(\frac{u}{2})\cosh(\frac{v}{2})})^2
\end{equation}
and so on. Now,
\begin{align}
    \bra{u} G^{-1}_T M \ket{v} &= \int d u' \; \bra{u} G^{-1}_T \ket{u'} \bra{u'} M \ket{v} \nn&= \delta(u-v) + \eta \bra{u} G_{T}^{-1} M_1 \ket{v} + \frac{\eta^2}{2} \bra{u} G_{T}^{-1} M_2 \ket{v} + \order{\eta^3}.
\end{align}
Therefore,
\begin{align}
    \log \det (G^{-1}_T M) &= \tr \log (G^{-1}_T M)\nn &= \eta \tr(G_T^{-1} M_1) + \frac{\eta^2}{2}\qty(
    \tr (G^{-1}_T M_2) - \tr\qty[(G_T^{-1}M_1)^2]
    )+\order{\eta^3}.
\end{align}
We now compute the regularized second R\'enyi-Shannon entropy. The idea is the same as before, except the volume of the hyberbolic space is different. If we assume $x \in (a+\epsilon, b-\epsilon)$, then the volume in the $u$ space is $2 \log(\frac{b-a}{\epsilon}) + \log(1-\eta)$. The same applies for the other interval. So, the second R\'enyi-Shannon entropy for $(-b,-a) \cup (a,b) \equiv A_- \cup A_+$ is
\begin{align}
    \sh_2 (\mathcal{O}_{\Delta=1}, A_- \cup A_+) &= \frac{1}{4} \qty(1-\frac{1}{2})^2 2 \times \qty[2 \log(\frac{b-a}{\epsilon}) + \log(1-\eta)] \nn & \quad +  \frac{\eta}2 \tr(G_T^{-1} M_1) + \frac{\eta^2}{4}\qty(
    \tr (G^{-1}_T M_2) - \tr[(G_T^{-1}M_1)^2]
    )+\order{\eta^3}\nn
    &= \frac{1}{4}\log(\frac{b-a}{\epsilon}) + \frac\eta 2 \qty(\tr(G_T^{-1} M_1) - \frac{1}{4}) \nn & \qquad \qquad+ \frac{\eta^2}{4}\qty(
    \tr (G^{-1}_T M_2) - \tr\qty[(G_T^{-1}M_1)^2]
    - \frac{1}{4})+\order{\eta^3}.\nn
\end{align}
When computing the measurement mutual information $I_2(A_-,A_+;\Delta=1)$, the first term above will cancel against the sum of the Shannon entropies of the individual intervals. The trace for the linear order in $\eta$ can be calculated
\begin{align}
    \tr(G_T^{-1} M_1) &= - \int d u \int d u' \int \frac{d k}{2 \pi} \; e^{i k (u-u')} \qty(-C_1 k \pi \coth(k \pi))^{-1} \frac{C_1}{8 \cosh^2(\frac{u}{2})\cosh^2(\frac{u'}{2})}  \nn
    &=  \int \frac{d k }{2 \pi} \; \frac{\tanh(k \pi)}{k \pi} \frac{1}{8}\qty(\frac{4 \pi k}{ \sinh(k \pi)})^2 = \frac{1}{4}.
\end{align}
As anticipated, the linear term vanishes, and the mutual information is of order $\eta^2$. We expect this to be true for any primary operator,
\begin{equation}
    I_2(\mathcal{O}; A_+, A_-) \propto \eta^{2 \Delta}.
\end{equation}
The larger $\Delta$ is, the faster the mutual information goes to zero for small $\eta$. Similar results are known for the mutual information of entanglement entropy, where the mutual information of entanglement entropy goes as $\eta^{2 \Delta_{\text{min}}}$, where $\Delta_{\text{min}}$ is the smallest scaling dimension of the primary operator that contributes to the decomposition of the sewing field in the replica CFT \cite{Cardy_2013}. 

\section{Discussion}
\noindent To summarize our results, we derived Shannon entropy and mutual information of measurement for the measurement of conformal primaries in two dimensional CFTs that are approximately at a Gaussian fixed point. Our derivation was based on the formula \eqref{eq:shannon-formula} for Shannon entropy of measurement. 
We have analytical control over \eqref{eq:shannon-formula} for CFTs in the large $N$ limit. For such theories in $d=2$ dimensions, and a {scalar} primary operator with scaling dimension $\Delta$, we derived the mutual information of measurement between two intervals that share a boundary to be
\begin{equation}
    I_n (\mathcal{O}; A, B) = \frac{1}{2}\qty(\Delta - \frac{1}{2})^2 \log(\frac{1}{1-\eta}), \quad \forall n \geq 1
\end{equation}
where $\eta$ is the cross ratio for the two intervals.
 We performed the derivation in two ways, one using Szeg\H{o} determinants, and the other using the Casini-Huerta-Myers (CHM) uniformization map that maps the correlators in vacuum in the original theory to thermal correlators. We note that this result is similar to that of R\'enyi mutual information  \cite{Calabrese_2009, Furukawa_2009, De_Nobili_2015, flam}, with the dependence on the CFT central charge replaced by a dependence on the conformal dimension of the operator which is measured.

The order of operations of measuring the field configurations of $\mathcal{O}$ and partitioning the space into intervals $A$ and $B$ is important. If we partition the space first, the above result will not hold. For example, measuring $\partial \phi$ has more information compared to measuring $\phi$ only if we measure first, and partition after. This way, $\partial \phi$ at the boundary of $A$ and $B$ shares information between the intervals. If we partition first and then measure $\partial \phi$ in points in $A$ and $B$ separately, this sharing is lost, and we cannot apply the above formula.

The idea that information is shared between 
the shared boundary between the intervals is strengthened by our calculation for mutual information for disjoint intervals, by a two-to-one map that generalizes the CHM map. The mutual information again depends on the cross ratio as 
\begin{equation}
    I_2 (\mathcal{O}; A, B) \propto \eta^{2 \Delta}.
\end{equation}
As expected, the mutual information vanishes as the intervals are separated farther, and it vanishes faster for larger scaling dimensions. This result is also similar to the R\'enyi mutual information  between disjoint intervals \cite{Cardy_2013}. 

There are a few immediate extensions of the current work which we leave for the future. Among them is extending this work to higher dimensional CFTs. In particular, one should be able to use the CHM map for  spherical regions to extract the $n$th R\'enyi-Shannon entropy. For measurement mutual information we need to generalize  to nested spherical regions that share a boundary: a sphere and its adjacent spherical shell seem the simplest choice. 
  Second, the CHM map for a single spherical region also yields a holographic bulk, the topological AdS black hole \cite{Casini:2011kv}. This can be leveraged if we want to compute corrections to mutual information in $1/N$.
  Though in this work we discussed mostly scalar operators, extensions to other primaries  should also be tractable in a similar fashion.

In this paper we explicitly worked out the mutual information for the two limits $\eta \to 1$ and $\eta \to 0$. It would be interesting to look for an all $\eta$ result, at least for special cases of CFTs. In the entanglement Rényi mutual information literature, finite $\eta$ results are only known for special cases of CFTs, such as CFTs with holographic duals \cite{Headrick:2010zt}, compact bosons and Ising CFT \cite{Calabrese_2009, Calabrese_2011}.  

While our results are valid for large $N$ CFTs, it seems likely that the mutual information of measurement is regularization independent, and the coefficient in front of the logarithm is universal. %for all CFTs. 
It is unlikely that we can apply the results derived in this article to CFTs for which higher point correlators cannot be neglected. However, on physical grounds, the coefficient can only depend on the dimension of the operator being measured, and possibly the CFT central charge. It would be interesting to study exactly what universal data is contained in the mutual information of measurement in general. A first piece of the puzzle can be found by deriving the leading $1/N$ correction. We leave this for future work. 

\section*{Acknowledgments}
I.K. would like to thank Y. Meurice and M.A. Rajabpour for useful discussions, as well as the Max Planck Institute MPI-PKS, Dresden for hospitality.

\appendix
\section{SWAP Formula for Entanglement Entropy}\label{app:SWAP}

In this appendix, we will elucidate the SWAP formula to calculate the second R\'enyi entanglement entropy \cite{klich2024swaptransposedisplacementsstabilizer} for any field theory. We will verify the formula explicitly for a simple non-interacting $2d$--CFT.

Define the SWAP operator on the direct product of the Hilbert space $\mathcal{H}$ with itself as
\begin{equation}
\swap : \mathcal{H} \otimes \mathcal{H} \to \mathcal{H} \otimes \mathcal{H}
\quad \text{with} \quad
\swap \ket{\psi,\phi} = \ket{\phi,\psi} \, .
\end{equation}
The SWAP operator that acts only on the region $A$, denoted by $\swap_A$, is simply
\begin{equation}
\swap_A = \swap\big|_{\mathcal{H}_A} \otimes \mathds{1}\big|_{\mathcal{H}_{A^c}} \, .
\end{equation}
It is well known that the second R\'enyi entantlement entropy for region $A$ for a pure state $\ket{\psi}$ can be written as
\begin{equation}
S_2(A) = - \log \qty( \bra{\psi,\psi} \swap_A \ket{\psi,\psi} ) \, .
\end{equation}
In fact, we can do a similar thing for any integer $n$, by having $n$ copies of the state $\psi$ and cyclically applying the SWAP operator. 

\par 
The SWAP operator can be re-written in a different way, using the displacement operator \cite{klich2024swaptransposedisplacementsstabilizer}. The displacement operator is defined in terms of phase space variables as  
\begin{equation}
{D}(q,p) = \exp\qty( i ( q \hat{Q} - p \hat{P} ) ) \, \qwhere q,p \in \reals.
\end{equation}
$\hat{Q}$ and $\hat{P}$ are the coordinate and the corresponding canonical momentum operators.
We can prove \cite{klich2024swaptransposedisplacementsstabilizer}
\begin{equation}
\swap
=
\int \frac{dq \, dp}{2\pi}
\;
{D}^\dagger(q,p) \otimes {D}(q,p) \, .
\end{equation}
Now, the SWAP operator that acts only on region $A$ is
\begin{equation}
\swap_A
=
\int \frac{dq \, dp}{2\pi}
\;
\qty( {D}^\dagger(q,p)\big|_A \otimes \mathds{1}_{A^c} ) \otimes
\qty( {D}(q,p)\big|_A \otimes \mathds{1}_{A^c} ) \, .
\end{equation}
With this, we can write a simple formula for the second R\'enyi entanglement entropy:
\begin{align}
S_2(A)
&=
- \log \qty(
\int \frac{dp \, dq}{2\pi}
\;
\abs{
\bra{\psi}
{D}(q,p)\big|_A \otimes \mathds{1}_{A^c}
\ket{\psi}
}^2
) \, .
\end{align}
Now, we apply the above formula to the case of a quantum field theory, where $\ket{\psi}$ is a many-body state. In this case, we will need to insert a SWAP operator for each one-body state. Therefore, 
\begin{equation}
\label{eq-s2formula}
S_2(A)
=
- \log \qty(
\int_A \mathcal{D}q(\vec x)\, \mathcal{D}p(\vec x)
\;
\abs{
\bra{\psi}
{D}[q(\vec x),p(\vec x)]
\ket{\psi}
}^2
) .
\end{equation}
with $p(\vec x)$ and $q(\vec x)$ vanishing outside $A$.

 We will calculate the entanglement entropy for a free scalar CFT and show that it matches the known result.
We pick $\ket{\psi}$ to be the vacuum of the theory. We will be general for now, and work in $d$ (Euclidean) dimensions. 
To compute the entanglement entropy, we first need to compute
\begin{equation}
    E[p, q] := \expval{\exp(i\int_{A}^{}d^{d-1}\vec x \; (q(\vec x)\hat \phi(0, \vec x) - p(\vec x)\hat \Pi(0, \vec x)))}
\end{equation}
This can be evaluated using the phase space (Euclidean) path integral, and is straightforward,
\begin{equation}
    \abs{E[p,q]}^2 = \exp(-\int_{A}^{}d^{d-1}\vec x d^{d-1}\vec y \; \qty[p(\vec x)F(\vec x-\vec y)p(\vec y) + q(\vec x)G(\vec x-\vec y)q(\vec y)] )
\end{equation}
where,
\begin{equation}
   \expval{\phi(\tau, \vec x), \phi(\tau, \vec y)}  = G(\vec x-\vec y)= \int_{\reals^{d-1}}^{} \frac{d^{d-1}\vec{k}}{(2\pi)^{d-1}} e^{-i\vec {k} \cdot(\vec {x} - \vec  y)} \frac{1}{2\sqrt{\vec  k^2 + m^2}}
\end{equation}
and,
\begin{equation}
    \expval{\pi(\tau, \vec x), \pi(\tau, \vec y)} = F(\vec x - \vec y) = \int_{\reals^{d-1}}^{} \frac{d^{d-1}\vec {k}}{(2\pi)^{d-1}} e^{-i\vec {k} \cdot(\vec {x} - \vec y)} \frac{\sqrt{\vec  k^2 + m^2}}{2}.
\end{equation}
We need to perform yet another path integral over \(p(\vec  x)\) and \(q(\vec  x)\). 
The integrals are Gaussian, and the results are functional determinants\footnote{We will work with the convention that $\Fint{\phi} \exp[- \frac{1}{2} \int d x d y \; \phi(x) M(x,y) \phi(y)] = [\det(M)]^{-1/2}$. }, with the matrices $F$ and $G$ restricted to the domain \(A \subset \reals^{d-1}\)
\begin{align}\label{eq-renyi-freefield-formal}
    S_2(A)
    &= \frac{1}{2}\qty[\log\qty(\textnormal{det}_A(2{{F}})) + \log\qty(\textnormal{det}_A(2{{G}}))]
\end{align}

We will specialize to \(d=2\) now. The spatial region will be taken to be the interval \(A = \qty[-\frac{\ell}{2}, \frac{\ell}{2}]\). We will find the asymptotic value of the entropy as \(\ell \to \infty\). We will make use of the Szeg\H{o} theorem described in Appendix \ref{app:Szego}. We use the discrete version of $G(x)$, given by 
\begin{equation}
    G_n = \int_{-\pi}^{\pi}\frac{d\theta}{2\pi} \frac{e^{-in\theta}}{2\sqrt{\sin^2(\theta) + \mu^2}}.
\end{equation}
The leading term in the Szeg\H{o} determinant cancels between ${F}$ and ${G}$, and the sub-leading terms add up. 
We get the second R\'enyi entropy to be:
\begin{equation}
    S_2(A) = \frac{1}{2}\log\qty(e^{\frac{1}{2}\log\qty(\frac{\ell}{\epsilon})}) = \frac{1}{4}\log\qty(\frac{\ell }{\epsilon})
\end{equation}
This is a known result, see for example equation (1.5) of \cite{Hung_2011}. 

\section{Szeg\H{o} Theorem and Toeplitz Determinants}\label{app:Szego}
 We will first define all the necessary objects and state the theorem.
Let \(\zeta_n \in \reals\) for \(n \in\ints\). Then, a Toeplitz matrix \(T_N \) is an $N \times N$ matrix with elements,
\begin{equation}\label{eq-def-Toeplitz}
    [T_N(\zeta)]_{mn} = \zeta_{m-n}
\end{equation}
 To state Szeg\H{o}'s theorem, we will consider smooth periodic functions on the circle \(S^1\), 
\begin{equation}
    \zeta : S^1 \to \reals \quad \textnormal{with} \quad \zeta(\theta + 2\pi) = \zeta(\theta)
\end{equation}
and define \(\zeta_n\)'s as,
\begin{equation}
    \zeta_n = \int_{-\pi}^{\pi}\frac{d\theta}{2\pi}\; e^{-in\theta}\zeta(\theta)
\end{equation}
We will then define the Toeplitz matrix as in \eqref{eq-def-Toeplitz}. Then Szeg\H{o}'s theorem states the following \cite{strongszego, deift2012toeplitzmatricestoeplitzdeterminants, simon2004sharpformstrongszego}\footnote{Check theorem 3 in \cite{deift2012toeplitzmatricestoeplitzdeterminants} or theorem 1.1 of \cite{simon2004sharpformstrongszego}.},\\
\noindent \textbf{Theorem(Szeg\H{o}) : }The determinant of the Toeplitz matrix in \eqref{eq-def-Toeplitz} has the asymptotic form,
\begin{equation}
    \lim_{N \to \infty} \det(T_N(\zeta)) = \lim_{N \to \infty} e^{N(\log \zeta)_0}e^{E_N(\zeta)},
\end{equation}
where,
\begin{equation}
    \qty(\log\zeta)_n \equiv \int_{-\pi}^{\pi}\frac{d\theta}{2\pi}\; e^{-in\theta} \log(\zeta(\theta)),
\end{equation}
and
\begin{equation}
    E_N(\zeta) = \sum_{n=1}^N n \abs{(\log\zeta)_n}^2
\end{equation}
Many proofs of the theorem are known, see for example \cite{simon2005orthogonal1}. A physically intuitive derivation of the theorem is given in \cite{Klich:2026mlr}.
For the purposes of this article, we consider periodic functions on  \(S^1\) given by,
\begin{equation}
    G(\theta) = \qty(\sin^2(\theta) + \mu^2)^{\Delta - \tfrac{1}{2}}, \quad \Delta \geq 0
\end{equation}
We construct a Toeplitz matrix out of the Fourier coefficients of this function. In the limit of $N \to \infty$, the leading term in the determinant is,
\begin{equation}
    N [\log(G)]_0 = N \qty(\Delta - \frac{1}{2}) \int_{-\pi}^{\pi}\frac{d\theta}{2\pi} \log(\sin^2(\theta) + \mu^2)
\end{equation}
For the sub-leading term, we need $[\log(G)]_n$ for $n>0$. We can do integration by parts,
\begin{align}
    \text{for } n>0: [\log(G)]_n &= \qty(\Delta - \frac{1}{2})\int_{-\pi}^{\pi}\frac{d\theta}{2\pi} e^{-i n \theta} \log(\sin^2(\theta) + \mu^2) \nn
    &= \frac{i}{n}\qty(\Delta - \frac{1}{2})\int_{-\pi}^{\pi}\frac{d\theta}{2\pi} e^{-in\theta} \frac{\sin(2\theta)}{\sin^2(\theta) + \mu^2}
\end{align}
The boundary term from integrating by parts vanish. 
It will be advantageous to go to continuum now to perform the integrals. Define $k = \frac{\theta}{\epsilon}$, 
\begin{align}
    (\log G)_n &= \frac{i}{n}\qty(\Delta - \frac{1}{2})\int_{-\pi/\epsilon}^{\pi/\epsilon} \frac{\epsilon dk}{2\pi} e^{-ik (n \epsilon)} \frac{\sin(2\epsilon k)}{\sin^2(\epsilon k) + \mu^2} \nn &\underset{\epsilon\to 0}{\approx} \frac{i}{n}\qty(\Delta - \frac{1}{2})\int_{-\infty}^{\infty} \frac{dk}{2\pi} e^{-ik(n \epsilon)}\frac{2k}{k^2 + (\mu/\epsilon)^2}\nn&= \frac{1}{ n}\qty(\Delta - \frac{1}{2}) e^{-\mu (n \epsilon)} 
\end{align}
This gives us,
\begin{align}
\sum_{n=1}^{N}
n\abs{[\log(G)]_n}^2
&=
\left(\Delta - \frac{1}{2}\right)^2
\sum_{n=1}^{N}
\frac{e^{-2\mu\epsilon n}}{n}.
\end{align}
Now we take the limit \(\mu\to0\) and approximate

\[
\sum_{n=1}^{N}
\rightarrow
\int_1^N dn ,
\]
which yields the subleading term
\begin{equation}
    E_N = \left(\Delta - \frac{1}{2}\right)^2 \log{N}.
\end{equation}
We therefore have
\begin{equation}\label{eq:szego-general-result}
    \log \det(T_N(G)) = N \qty(\Delta - \frac{1}{2}) \int_{-\pi}^{\pi}\frac{d\theta}{2\pi} \log(\sin^2(\theta) + \mu^2) + \left(\Delta - \frac{1}{2}\right)^2 \log{N}.
\end{equation}

\section{$\Delta = 1/2$}\label{app:Delta-half}
For the $\Delta = 1/2$ Green's function of a scalar CFT, the Fourier transform is
\begin{equation}
\mathcal{G}_{1/2}(k) = -2\left(\gamma + \log(\epsilon |k|)\right).
\end{equation}
The constant $\gamma$ can be absorbed into the cutoff $\epsilon$, so we will drop it. When finding the sub-leading term in Szeg\H{o}, we will need
\begin{equation}
[\log \mathcal{G}_{1/2}]_x = \int dk\; e^{-ikx} \log\big(\log(\epsilon |k|)\big).
\end{equation}
This term is going to be cutoff dependent. We can do the integration by parts trick,
\begin{equation}
x \neq 0: \qquad [\log \mathcal{G}_{1/2}]_x = \frac{i}{x} \int_{\reals} dk\; e^{-ikx} \frac{1}{k \log(\epsilon |k|)}.
\end{equation}
We will find the behavior of the above in the large $x$ limit. Under a variable change $k = \xi/x$, we end up with
\begin{equation}
[\log \mathcal{G}_{1/2}]_x = \frac{i}{x} \int_{\reals} d\xi\; \frac{1}{\xi \left(\log|\xi| - \log(|x|/\epsilon)\right)}.
\end{equation}
For large $x$, therefore,
\begin{equation}
[\log \mathcal{G}_{1/2}]_x \sim \frac{1}{x \log(|x|)}.
\end{equation}
Therefore, the strong Szeg\H{o} sum,
\begin{equation}
\int^{\infty} dx\; x \left|\frac{1}{x \log(|x|)}\right|^2,
\end{equation}
is convergent and is a number independent of the log of the size of the interval, making it an $\order{1}$ contribution to the mutual information.

\section{The Increase in Mutual Information vs Information Processing Inequality}\label{app-information-processing}
The result in \eqref{eq:I2-O} for the mutual information of the adjacent intervals looks like it contradicts the information processing inequality. In this section, we will describe how to resolve this apparent tension. For simplicity, let us go back to a discrete lattice. Let $X$ denote the field that we measure, and the measurement outcomes looks,
\begin{equation}
    X = (x_1, \dots, x_L), \quad L = \ell_A + \ell_B
\end{equation}
We split this into two intervals $A$ and $B$, and consider random variables $X_A = (x_1, \dots, x_{\ell_A})$ and $X_B = (x_{\ell_A + 1}, \dots, x_L)$. Denote the mutual information between these two random variables to be $I(X_A; X_B)$. We now perform local information processing on both intervals $A$ and $B$ as,
\begin{equation}
    Z_A = f_A(X_A), \quad \text{and} \quad Z_B = f_B(X_B)
\end{equation}
The map $f$ can be deterministic or stochastic. Then, the information processing inequality states~\cite{Cover2006},
\begin{equation}
    I(Z_A; Z_B) \leq I(X_A; X_B)
\end{equation}
That is, local processing of information can only destroy information shared between the two random variables. 

The apparent tension with our result is the following: measuring the field configurations and taking a derivative is just information processing. How can the mutual information then increase? To resolve this apparent paradox, consider a particular processing, the discrete derivative. If we first partition the the system into two, and then process the information, we will have the two new random variables,
\begin{align}
    Y_A &= D_A X_A
=
(x_2-x_1,\ldots,x_{\ell_A}-x_{\ell_A-1}),
\nonumber \\
Y_B &= D_B X_B
=
(x_{\ell_A+2}-x_{\ell_A+1},
   \ldots,
   x_{\ell_A+\ell_B}-x_{\ell_A+\ell_B-1}).
\end{align}
where $D_A,D_B$ are local discrete derivatives, and therefore,
\begin{equation}
    I(Y_A; Y_B) \leq I(X_A; X_B)
\end{equation}
The resolution to the puzzle is the following: the quantity we computed is not the same as $I(Y_A; Y_B)$. Consider,
\begin{align}
    Y = D X &= (y_1, \dots, y_{\ell_A - 1}, y_{\ell_A}, y_{\ell_A + 1}, \dots, y_{L-1}) \nn &= (x_2-x_1, \dots, x_{\ell_A} - x_{\ell_A-1}, x_{\ell_A + 1} - x_{\ell_A},x_{\ell_A + 2} - x_{\ell_A+1}, \dots, x_{L} - x_{L-1} )
\end{align}
We see that if we now partition $Y$ into two, There is an extra bond, $y_\partial:= y_{\ell_A} = x_{\ell_A + 1} - x_{\ell_A}$ which gets assigned to one of the partitions. For
example, one may assign the interface variable to the left side,
\begin{align}
    \widetilde Y_A
    &=
    (y_1,\ldots,y_{\ell_A-1},y_\partial),\\
    \widetilde Y_B
    &=
    (y_{\ell_A+1},\ldots,y_{L-1}).
\end{align}
If we first partition space, and measure the differences in each of the subspaces, then the boundary variable is never measured. In short, $Y = \widetilde{Y}_A \oplus \widetilde{Y}_B \neq Y_A \oplus Y_B$. The transformation from
$(X_A,X_B)$ to $(\widetilde Y_A,\widetilde Y_B)$ is not a pair of local
data-processing maps, and consequently the information-processing inequality
does not apply. 
Thus, what resolves the apparent contradiction is the order of operations. Partition first and then measure is not the same as measure first and then partition. In our set up of the problem, we are doing the latter.

We will now explicitly demonstrate the increase in information for our specific example. Let $X$ be a Gaussian field with covariance in the Fourier space\footnote{The continuum limit is recovered in the limit $\theta \to 0$.},
\begin{equation}
    g_\Delta(\theta) \propto \qty(4 \sin^2(\theta/2))^{\Delta - \tfrac{1}{2}}
\end{equation}
Taking the discrete derivative will multiply the covariance matrix in Fourier space by $\abs{e^{i\theta}-1}^2=4\sin^2(\theta/2)$, increasing $\Delta \to \Delta + 1$ in the covariance of $Y$. Therefore, measuring $Y$'s everywhere first, and then partitioning will see an increase in mutual information. We will now show that this increase is purely coming from the bound at the boundary. The covariance matrix of $Y$ has the form,
\begin{equation}
    C_Y = \begin{pmatrix}
C_{AA} & C_{A\partial} & C_{AB}\\
C_{A\partial}^T & C_{\partial \partial} & C_{B\partial}^T\\
C_{AB}^T & C_{B\partial} & C_{BB}
\end{pmatrix}.
\end{equation}
Here $C_{AA}$ and $C_{BB}$ are the covariance matrices of the derivatives lying
strictly inside $A$ and $B$, respectively, while the middle row and column
correspond to the correlations involving the  interface derivative $y_\partial$.

If instead we first partition the system and then differentiate
independently inside $A$ and $B$, the interface variable $y_\partial$ is
never measured.  The measured variables are therefore only $(Y_A,Y_B)$,
whose covariance matrix is
\begin{equation}
C_{(Y_A,Y_B)}
=
\begin{pmatrix}
C_{AA} & C_{AB}\\
C_{AB}^T & C_{BB}
\end{pmatrix}.
\label{eq:local-derivative-covariance}
\end{equation}
The above covariance matrix is obtained by removing the row and column involving the boundary correlators. The standard principal-minor identity
gives
\begin{equation}
    \det C_{(Y_A,Y_B)}
    =
    \det C_{Y}\,
    \bigl(C_{Y}^{-1}\bigr)_{\ell_A\ell_A}.
    \label{eq:principal-minor-interface}
\end{equation}
Let us denote the mutual information for ``first measure then partition" as $I(Y; A, B)$ and the mutual information for ``first partition then measure" by $I(Y_A; Y_B)$. For measuring first and partitioning after, let us arbitrarily assign the boundary measurement to $A$. Denote,
\begin{equation}
    \widetilde{C}_{AA} = \mqty(C_{AA} & C_{A\partial}\\ C^T_{A\partial} & C_{\partial \partial})
\end{equation}
Then,
\begin{equation}
    I(Y; A, B) = \frac{1}{2}\log\qty(\frac{\det(\widetilde{C}_{AA}) \det(C_{BB})}{\det C_Y}) = \frac{1}{2}\qty(\Delta + \frac{1}{2})^2 \log(\frac{\ell_A \ell_B}{\ell_A + \ell_B})
\end{equation}
The reason once again is that $C_Y$ in Fourier space is obtained by multiplying $C_X(k)$ by $4 \sin^2(\theta/2)$, increasing $\Delta$ by one. In contrast for partitioning first and measuring after,
\begin{equation}
    I(Y_A; Y_B) = \frac{1}{2}\log\qty(\frac{\det(C_{AA}) \det(C_{BB})}{\det C_{(Y_A, Y_B)}}) = I(Y; A, B) - \frac{1}{2}\log(\frac{\det(\widetilde{C}_{AA})}{\det(C_{AA})})- \frac{1}{2} \log \qty(\bigl(C_{Y}^{-1}\bigr)_{\ell_A\ell_A})
\end{equation}
The correction coming from the ratio of the determinants of $\widetilde{C}_{AA}$ and $C_{AA}$ is of $\order{1}$, since both are Toeplitz determinants with the same symbol, with sizes differing by only one. This is a sub-leading contribution in large interval sizes to the mutual information. 
\old
The correction from the single matrix element at the interface can be estimated as follows\footnote{A more rigorous estimate can be done with asymptotics of a Toeplitz matrix with Fisher-Hartwig zero of the type $|1-e^{i\theta}|^{2\alpha}h(e^{i\theta})$, with smooth and strictly positive function $h$. }: for $j$ away from the edges of the interval $[1,L]$, the diagonal entries of the inverse matrix are,
\begin{equation}
    (C_{Y}^{-1}\bigr)_{j j} \sim \int_{-\pi}^\pi \frac{d \theta}{2 \pi}\; g_{\Delta+1}^{-1}(\theta)
\end{equation}
due to the zero of $g_{\Delta+1}$ at $\theta = 0$, the integral will be dominated by the contribution from this pole. The divergence is cut-off at $\theta_{\text{min}} \sim \frac{\pi}{L}$. Therefore, we get,
\begin{equation}
     (C_{Y}^{-1}\bigr)_{j j} \sim \int_{\pi/L}^1 \frac{d \theta}{\abs{\theta}^{2\qty(\Delta + 1/2)}} \sim L^{2\Delta} + \order{1}
\end{equation}
Thus,
\begin{equation}
   \log \bigl(C_{Y}^{-1}\bigr)_{\ell_A\ell_A} \sim 2 \Delta \log L + \order{1}
\end{equation}
For $\ell_A/\ell_B$ fixed, we have,
\begin{equation}
    \log\frac{\ell_A\ell_B}{\ell_A+\ell_B}
    =
    \log L+O(1),
\end{equation}
Thus, the correction term gives us,
\begin{equation}
    I(Y_A; Y_B) \approx \qty[\frac{1}{2} \qty(\Delta + \frac{1}{2})^2 - \Delta] \log(\frac{\ell_A \ell_B}{\ell_A + \ell_B}) = I(X_A; X_B)
\end{equation}
We explicitly see that the extra information is coming from the correlation at the boundary. When we are only doing information processing, the boundary value of $Y$ is never measured, and taking away this extra information restores the information processing inequality. 

\section{Exact Calculations for Simple Models}
\subsection{Coupled Oscillators}\label{app:oscillator}
Consider two coupled oscillators with the Hamiltonian
\begin{equation}
H = \frac{p_1^2}{2m} + \frac{p_2^2}{2m} + \frac{1}{2} m \omega^2 (x_1^2 + x_2^2) + \frac{1}{2} m \omega^2 \lambda (x_1 - x_2)^2.
\end{equation}
We can use this as a prototype for a discretized free scalar field theory.
We diagonalize the Hamiltonian by a simple change of variables
\begin{equation}
x_{\pm} = \frac{x_1 \pm x_2}{\sqrt{2}},
\end{equation}
and obtain
\begin{equation}
H = \frac{p_+^2}{2m} + \frac{p_-^2}{2m} + \frac{1}{2} m \omega^2 (x_+^2 + x_-^2) + m \omega^2 \lambda x_-^2.
\end{equation}
The ground state wave function is quite simple,
\begin{equation}
\psi_0(x_+, x_-) = \sqrt{\frac{m \omega}{\pi}} \xi^{1/4} \exp\!\left(-\frac{m \omega}{2} (x_+^2 + \xi\, x_-^2)\right),
\end{equation}
where $\xi = \sqrt{1 + 2\lambda}$. We can now go back to the original basis of $x_{1,2}$, and find the density matrix.

\par  To find the measurement entropy of  position outcomes, we need the probability distribution,
\begin{equation}
P_{12}(x_1, x_2) = \left|\psi_0(x_1, x_2)\right|^2.
\end{equation}
From this, we can find the partial probability to measure only $x_1$ by tracing over $x_2$,
\begin{equation}
P_1(x_1) = \int dx_2 \; P_{12}(x_1, x_2).
\end{equation}
The second R\'enyi-Shannon entropy of measuring the first oscillator is then,
\begin{equation}
\mathrm{Sh}_2(1) = -\log\!\left[\int dx_1 \; P_1^2(x_1)\right] = \frac{1}{2}\log\!\left(\frac{\pi}{m\omega} \frac{1+\xi}{\xi}\right).
\end{equation}
We can also find the full entropy,
\begin{align}
\mathrm{Sh}_2(1 \cup 2) = \int dx_1\, dx_2 \; P_{12}^2(x_1, x_2) 
&= \log\!\left(\frac{\pi}{m\omega} \frac{2}{\sqrt{\xi}}\right).
\end{align}
We compute the mutual information of measurement,
\begin{align}
I_2(1;2) &= \mathrm{Sh}_2(1) + \mathrm{Sh}_2(2) - \mathrm{Sh}_2(1 \cup 2) 
= \log\!\left(\frac{1+\xi}{2\sqrt{\xi}}\right).
\end{align}
We re-derive the above results using the formula \eqref{eq:shannon-formula} as a simple check. 
We need to evaluate,
\begin{equation}
e^{-\mathrm{Sh}_2(1)} = \int \frac{dq_1}{\sqrt{2}\,\pi} \left|\left\langle e^{i q_1 \hat{x}_1} \right\rangle\right|^2.
\end{equation}
The normalization of the integral is picked to match the direct result. Since we have a Gaussian theory,
\begin{equation}
\left\langle e^{i q_1 \hat{x}_1} \right\rangle = e^{-q_1^2 \langle \hat{x}_1^2 \rangle}
\end{equation}
giving us,
\begin{equation}
e^{-\mathrm{Sh}_2(1)} = \frac{1}{\sqrt{4\pi}} \frac{1}{\sqrt{\langle \hat{x}_1^2 \rangle}}\label{Sh21}
\end{equation}
we have,
\begin{align}
\langle \hat{x}_1^2 \rangle = \frac{1}{2}\left[\langle \hat{x}_+^2 \rangle + \langle \hat{x}_-^2 \rangle\right] 
= \frac{1}{2}\left[\frac{1}{2m\omega} + \frac{1}{2m\xi\omega}\right] = \frac{1}{4m\omega}\frac{\xi+1}{\xi}
\end{align}
Substituting into \eqref{Sh21}, we get
\begin{equation}
\mathrm{Sh}_2(1) = \frac{1} {2}\log\!\left(\frac{\pi}{m\omega}\frac{1+\xi}{\xi}\right).
\end{equation}
Next, to find the second R\'enyi-Shannon entropy for measuring the position of both particles, we evaluate
\begin{equation}
e^{-\mathrm{Sh}_2(1 \cup 2)} = \int \frac{dq_1\, dq_2}{2\pi^2} \left|\left\langle e^{i q_1 \hat{x}_1 + i q_2 \hat{x}_2} \right\rangle\right|^2.\label{sh212}
\end{equation}
Once again, we can go to the normal basis,
\begin{align}
\left\langle e^{i q_1 \hat{x}_1 + i q_2 \hat{x}_2} \right\rangle
&= \left\langle e^{\frac{i}{\sqrt{2}}\left[(q_1+q_2)\hat{x}_+ + (q_1-q_2)\hat{x}_-\right]} \right\rangle \nonumber \\
&= e^{-\frac{1}{2}(q_1+q_2)^2 \langle \hat{x}_+^2 \rangle} \, e^{-\frac{1}{2}(q_1-q_2)^2 \langle \hat{x}_-^2 \rangle} \nonumber \\
&= e^{-\frac{1}{4m\omega}\left[(q_1+q_2)^2 + (q_1-q_2)^2/\xi\right]}.
\end{align}
Substituting into \eqref{sh212}, and computing the $q$ integrals yields
\begin{equation}
e^{-\mathrm{Sh}_2(1 \cup 2)} = \frac{m\omega}{2\pi}\sqrt{\xi}.
\end{equation}
We have arrived at the same result as the direct calculation.
\subsection{Free Scalar Field}\label{app:freescalar}
We will calculate the mutual information of measurement for a free scalar field $\phi$. 
We first need the wave function(al) for the ground state. This can be expressed as the path integral
\begin{equation}
\Psi_0[\phi(x), \tau=0] = \int_{\chi(x,-\infty)=0}^{\chi(x,0)=\phi(x)} \mathcal{D}\chi \, 
\exp\Bigg[-\int dx \int_{-\infty}^{0} d\tau \, \frac{1}{2} \big(\dot{\chi}^2 + (\chi')^2 + m^2 \chi^2\big)\Bigg]
\label{eq:phi-path-integral}.
\end{equation}
To compute the integral, we can solve for the equations of motion, and find the on-shell action. 
\begin{equation}
\Psi_0[\phi(x)] = \mathcal{N} \exp\Bigg[-\frac{1}{2} \int dx dy \, \phi(x) \left( \int \frac{dk}{2\pi} e^{-i k (x-y)} \omega_k \right) \phi(y)\Bigg]
\label{eq:phi-wavefunction}
\end{equation}
The normalization constant is:
\begin{equation}
\mathcal{N} = \Big[\det \qty( \int \frac{dk}{2\pi} e^{-i k (x-y)} \omega_k ) \Big]^{1/4}
\label{eq:phi-normalization}
\end{equation}
Let $A \subset \reals$ be a spatial slice. We will trace over $A^c = \reals\setminus A$. We decompose the field according to:
\begin{equation}
\phi(x) = \phi_A(x) + \phi_{A^c}(x)\equiv \phi_+ + \phi_-
\label{eq:phi-decomposition},
\end{equation}
and introduce the notation
\begin{equation}
\int \frac{dk}{2\pi} e^{-i k (x-y)} \omega_k \equiv M_{xy}
\label{eq:M-xy}.
\end{equation}
Then,  schematically, using a notation similar to that of Callan and Wilczek \cite{Callan-Wilczek} where the integrals are suppressed, we have:
\begin{equation}
P[\phi] = \mathcal{N}^2 \exp\Big[- (\phi_+ \, \phi_-) 
\begin{pmatrix} M_{++} & M_{+-} \\ M_{-+} & M_{--} \end{pmatrix}
\begin{pmatrix} \phi_+ \\ \phi_- \end{pmatrix} \Big]
\label{eq:probability-matrix}
\end{equation}
Integrating $\phi_-$, we get:
\begin{equation}
P_+[\phi_+] = \mathcal{N}' \exp\Big[- \phi_+ M_{++} \Big( 1 - M_{++}^{-1} M_{+-} M_{--}^{-1} M_{-+} \Big) \phi_+ \Big]
\label{eq:phi-plus-probability}
\end{equation}
The inverses are inverses of the restricted matrix, not the restriction of $M^{-1}$. The normalization constant is:
\begin{equation}
\mathcal{N}' = \Big[ \det \Big(M_{++} \big( 1 - M_{++}^{-1} M_{+-} M_{--}^{-1} M_{-+} \big) \Big) \Big]^{1/2} \equiv \sqrt{\det(K_{++})}
\label{eq:phi-plus-normalization}
\end{equation}
We can square the probability and integrate over $\phi_+$ and find the second Rényi measurement entropy:
\begin{equation}
\sh_2 ({\phi}; A) = - \log \Big[ \frac{\det(K_{++})}{\sqrt{\det(2 K_{++})}} \Big] = -\frac{1}{2} \mathrm{tr} \log \frac{K_{++}}{2}
\label{eq:phi-second-renyi}
\end{equation}
Assuming $M_{--}$ and $M_{++} - M_{+-}(M_{--})^{-1} M_{-+}$ are invertible, the block inverse of $M$ has the $++$ component to be,
\begin{equation}
    (M^{-1})_{++} = (K_{++})^{-1},
\end{equation}
which yields
\begin{equation}
\sh_2({\phi}; A) = \frac{1}{2} \mathrm{tr} \log \big( 2 (M^{-1})_{++} \big),
\label{eq:phi-second-renyi-final}
\end{equation}
which is the same as what we would get for a scalar field from \eqref{eq:shannon-formula}. The determinant can be then computed for the massless case, using the Szeg\H{o} formula with $\Delta = 0$, and obtain:
\begin{equation}
I_2({\phi}; A, B) = \frac{1}{8} \log{\frac{1}{1-\eta}}.
\label{eq:mutual-information-massless}
\end{equation}

\subsection{Free Fermions }\label{app:freefermion}
Consider a free massless Dirac fermion in $1+1$ dimensions,
\begin{equation}
\mathcal S = \int d^2 x \, i \bar{\psi} \gamma^\mu \partial_\mu \psi
\label{eq:dirac-action}.
\end{equation}
We choose the Dirac matrices to be the Pauli matrices,
\begin{equation}
\gamma^1 = \sigma^1, \quad \gamma^2 = i \sigma^2
\label{eq:dirac-matrices}.
\end{equation}
The action separates into left-moving and right-moving sectors,
\begin{equation}
\mathcal S = \int d^2 x \, \Big( \psi_+^\dagger (\partial_\tau - \partial_x) \psi_+ + \psi_-^\dagger (\partial_\tau + \partial_x) \psi_- \Big)
\label{eq:left-right-action},
\end{equation}
where $\psi_\pm$ are eigenvectors of $\gamma^5 = \sigma^3$.
We will need the two-point functions at equal time:
\begin{equation}
\langle \psi_\pm(x,0) \psi_\pm^\dagger(0,0) \rangle = \frac{1}{2\pi x}
\label{eq:two-point}
\end{equation}
and the left and right movers are uncorrelated.
\par 
We calculate the mutual information of measurement of the charge density operator $\psi^\dagger\psi$. To this end, we need to evaluate:
\begin{equation}
\sh_2(\psi^\dagger \psi; A) = -\log \Bigg[ \int_A \mathcal{D}q(x) \, \Big| \Big\langle e^{i \int_A dx \, q(x) \psi^\dagger(x) \psi(x)} \Big\rangle \Big|^2 \Bigg]
\label{eq:shannon-entropy}
\end{equation}
The number density must be defined carefully to avoid UV issues. This is usually done via point splitting:
\begin{equation}
n(x) := \lim_{\epsilon \to 0} \Big[ \psi^\dagger(x+\epsilon) \psi(x) - \langle \psi^\dagger(\epsilon) \psi(0) \rangle \Big]
\label{eq:rho-point-splitting}
\end{equation}
by definition,
\begin{equation}
\langle n(x) \rangle = 0
\label{eq:rho-zero}
\end{equation}
All higher cumulants except the two-point to vanish. This is clear from the bosonized picture.
Therefore,
\begin{equation}
\Big\langle e^{i \int_A dx \, q(x) \psi^\dagger(x) \psi(x)} \Big\rangle = \exp\Big[ - \int_A dx dx' \, q(x) \langle n(x) n(x') \rangle q(x') \Big]
\label{eq:exponential-rho}
\end{equation}
Then, the second R\'enyi-Shannon entropy can be found using the Szeg\H{o} determinant. As
\begin{equation}
\langle n(x) n(0) \rangle = \frac{1}{\pi^2 x^2}
\label{eq:rho-two-point}
\end{equation}
We have $\Delta = 1$, hence the result:
\begin{equation}
I_2(n; A,B) =  \frac{1}{8}  \log {\frac{1}{1-\eta}}
\label{eq:shannon-final}
\end{equation}

\bibliography{biblio}

@article{Hung_2011,
   title={Holographic calculations of Rényi entropy},
   volume={2011},
   ISSN={1029-8479},
   url={http://dx.doi.org/10.1007/JHEP12(2011)047},
   DOI={10.1007/jhep12(2011)047},
   number={12},
   journal={Journal of High Energy Physics},
   publisher={Springer Science and Business Media LLC},
   author={Hung, Ling-Yan and Myers, Robert C. and Smolkin, Michael and Yale, Alexandre},
   year={2011},
   month=dec 
}

@article{Stephan2009,
  author  = {St{\'e}phan, Jean-Marie and Furukawa, Shunsuke
             and Misguich, Gr{\'e}goire and Pasquier, Vincent},
  title   = {{Shannon} and entanglement entropies of one- and
             two-dimensional critical wave functions},
  journal = {Phys. Rev. B},
  volume  = {80},
  pages   = {184421},
  year    = {2009},
  doi     = {10.1103/PhysRevB.80.184421}
}

@article{Stephan2011,
  author  = {St{\'e}phan, Jean-Marie and Misguich, Gr{\'e}goire
             and Pasquier, Vincent},
  title   = {Phase transition in the {R{\'e}nyi-Shannon} entropy
             of {Luttinger} liquids},
  journal = {Phys. Rev. B},
  volume  = {84},
  pages   = {195128},
  year    = {2011},
  doi     = {10.1103/PhysRevB.84.195128}
}

@article{AlcarazRajabpour2013,
  author  = {Alcaraz, F. C. and Rajabpour, M. A.},
  title   = {Universal behavior of the {Shannon} mutual information
             of critical quantum chains},
  journal = {Phys. Rev. Lett.},
  volume  = {111},
  pages   = {017201},
  year    = {2013},
  doi     = {10.1103/PhysRevLett.111.017201}
}

@article{Stephan2014,
  author  = {St{\'e}phan, Jean-Marie},
  title   = {{Shannon} and {R{\'e}nyi} mutual information
             in quantum critical spin chains},
  journal = {Phys. Rev. B},
  volume  = {90},
  pages   = {045424},
  year    = {2014},
  doi     = {10.1103/PhysRevB.90.045424}
}

@article{AlcarazRajabpour2014,
  author  = {Alcaraz, F. C. and Rajabpour, M. A.},
  title   = {Universal behavior of the {Shannon} and {R{\'e}nyi}
             mutual information of quantum critical chains},
  journal = {Phys. Rev. B},
  volume  = {90},
  pages   = {075132},
  year    = {2014},
  doi     = {10.1103/PhysRevB.90.075132}
}

@article{JinKorepin2004,
  author        = {Jin, B.-Q. and Korepin, V. E.},
  title         = {Quantum Spin Chain, {Toeplitz} Determinants
                   and the {Fisher--Hartwig} Conjecture},
  journal       = {Journal of Statistical Physics},
  volume        = {116},
  number        = {1--4},
  pages         = {79--95},
  year          = {2004},
  doi           = {10.1023/B:JOSS.0000037230.37166.42},
  eprint        = {quant-ph/0304108},
  archivePrefix = {arXiv}
}

@article{ItsJinKorepin2005,
  author        = {Its, A. R. and Jin, B.-Q. and Korepin, V. E.},
  title         = {Entanglement in the {XY} Spin Chain},
  journal       = {Journal of Physics A: Mathematical and General},
  volume        = {38},
  number        = {13},
  pages         = {2975--2990},
  year          = {2005},
  doi           = {10.1088/0305-4470/38/13/011},
  eprint        = {quant-ph/0409027},
  archivePrefix = {arXiv}
}

@article{AlcarazRajabpour2015,
  author  = {Alcaraz, F. C. and Rajabpour, M. A.},
  title   = {Generalized mutual information of quantum critical chains},
  journal = {Phys. Rev. B},
  volume  = {91},
  pages   = {155122},
  year    = {2015},
  doi     = {10.1103/PhysRevB.91.155122}
}

@article{Meurice2025,
  author  = {Meurice, Yannick},
  title   = {Lower bounds on entanglement entropy without twin copy},
  journal = {Phys. Rev. Research},
  volume  = {7},
  pages   = {L022023},
  year    = {2025},
  doi     = {10.1103/PhysRevResearch.7.L022023}
}

@article{Kaufman2025,
  author  = {Kaufman, Avi and Corona, James and Ozzello, Zane
             and Senseman, Blake and Asaduzzaman, Muhammad
             and Meurice, Yannick},
  title   = {Improved entanglement entropy estimates
             from filtered bitstring probabilities},
  journal = {Phys. Rev. A},
  volume  = {112},
  pages   = {032430},
  year    = {2025},
  doi     = {10.1103/lbv3-2z7r}
}

@article{Koyluoglu2026,
  author  = {K{\"o}yl{\"u}o{\u{g}}lu, N. U.
             and Majumder, Swarnadeep
             and Amico, Mirko
             and Mostame, Sarah
             and van den Berg, Ewout
             and Rajabpour, M. A.
             and Minev, Zlatko
             and Najafi, Khadijeh},
  title   = {Measuring central charge on a universal quantum processor},
  journal = {Nature Communications},
  volume  = {17},
  pages   = {305},
  year    = {2026},
  doi     = {10.1038/s41467-025-66775-9}
}

@article{klich2024swaptransposedisplacementsstabilizer,
      title={SWAP and Transpose by displacements, Stabilizer Renyi entropies for continuous variables and qudits and other applications}, 
      author={Israel Klich},
      year={2024},
      eprint={2408.15161},
      archivePrefix={arXiv},
      primaryClass={quant-ph},
      journal = {arXiv preprint}
}

@article{deift2012toeplitzmatricestoeplitzdeterminants,
author = {Deift, Percy and Its, Alexander and Krasovsky, Igor},
title = {Toeplitz Matrices and Toeplitz Determinants under the Impetus of the Ising Model: Some History and Some Recent Results},
journal = {Communications on Pure and Applied Mathematics},
volume = {66},
number = {9},
pages = {1360-1438},
doi = {https://doi.org/10.1002/cpa.21467},
url = {https://onlinelibrary.wiley.com/doi/abs/10.1002/cpa.21467},
eprint = {https://onlinelibrary.wiley.com/doi/pdf/10.1002/cpa.21467},
year = {2013}
}

@article{simon2004sharpformstrongszego,
      title={The sharp form of the strong Szego theorem}, 
      author={Barry Simon},
      year={2004},
      eprint={math/0402110},
      archivePrefix={arXiv},
      primaryClass={math.SP},
      journal = {arXiv preprint}
}

@article {strongszego,
    AUTHOR = {Szeg\"o, G.},
     TITLE = {On certain {H}ermitian forms associated with the {F}ourier
              series of a positive function},
   JOURNAL = {Comm. S\'em. Math. Univ. Lund [Medd. Lunds Univ. Mat. Sem.]},
  FJOURNAL = {Communications du S\'eminaire Math\'ematique de
              l'Universit\'e{} de Lund [Meddelanden fr\aa n{} Lunds
              Universitets Matematiska Seminarium]},
    VOLUME = {1952},
      YEAR = {1952},
     PAGES = {228--238},
      ISSN = {0373-5613},
   MRCLASS = {42.4X},
  MRNUMBER = {51961},
MRREVIEWER = {R.\ Salem},
}

@article{Callan-Wilczek,
    author = "Callan, Jr., Curtis G. and Wilczek, Frank",
    title = "{On geometric entropy}",
    eprint = "hep-th/9401072",
    archivePrefix = "arXiv",
    reportNumber = "IASSNS-HEP-93-87",
    doi = "10.1016/0370-2693(94)91007-3",
    journal = "Phys. Lett. B",
    volume = "333",
    pages = "55--61",
    year = "1994"
}

@article{Calabrese-Cardy,
    author = "Calabrese, Pasquale and Cardy, John L.",
    title = "{Entanglement entropy and quantum field theory}",
    eprint = "hep-th/0405152",
    archivePrefix = "arXiv",
    doi = "10.1088/1742-5468/2004/06/P06002",
    journal = "J. Stat. Mech.",
    volume = "0406",
    pages = "P06002",
    year = "2004"
}

@article{Kullback:1951zyt,
    author = "Kullback, S. and Leibler, R. A.",
    title = "{On Information and Sufficiency}",
    doi = "10.1214/aoms/1177729694",
    journal = "The Annals of Mathematical Statistics",
    volume = "22",
    number = "1",
    pages = "79--86",
    year = "1951"
}

@article{Freedman:1998tz,
  author = {Freedman, Daniel Z. and Mathur, Samir D. and Matusis, Arvind and Rastelli, Leonardo},
  title = {Correlation functions in the CFT(d)/AdS(d+1) correspondence},
  journal = {Nucl. Phys. B},
  volume = {546},
  pages = {96--118},
  year = {1999},
  doi = {10.1016/S0550-3213(99)00053-X},
  eprint = {hep-th/9804058},
  archivePrefix = {arXiv},
  primaryClass = {hep-th}
}

@article{Hartman_2008,
   title={Double-trace deformations, mixed boundary conditions and functional determinants in AdS/CFT},
   volume={2008},
   ISSN={1029-8479},
   url={http://dx.doi.org/10.1088/1126-6708/2008/01/019},
   DOI={10.1088/1126-6708/2008/01/019},
   number={01},
   journal={Journal of High Energy Physics},
   publisher={Springer Science and Business Media LLC},
   author={Hartman, Thomas and Rastelli, Leonardo},
   year={2008},
   month=jan, pages={019–019} }

@article{witten2002multitraceoperatorsboundaryconditions,
      title={Multi-Trace Operators, Boundary Conditions, And AdS/CFT Correspondence}, 
      author={Edward Witten},
      year={2002},
      eprint={hep-th/0112258},
      archivePrefix={arXiv},
      primaryClass={hep-th},
      journal = {arXiv preprint}
}

@article{JLMS,
   title={Relative entropy equals bulk relative entropy},
   volume={2016},
   ISSN={1029-8479},
   url={http://dx.doi.org/10.1007/JHEP06(2016)004},
   DOI={10.1007/jhep06(2016)004},
   number={6},
   journal={Journal of High Energy Physics},
   publisher={Springer Science and Business Media LLC},
   author={Jafferis, Daniel L. and Lewkowycz, Aitor and Maldacena, Juan and Suh, S. Josephine},
   year={2016},
   month=June }

@article{Casini:2011kv,
    author = "Casini, Horacio and Huerta, Marina and Myers, Robert C.",
    title = "{Towards a derivation of holographic entanglement entropy}",
    eprint = "1102.0440",
    archivePrefix = "arXiv",
    primaryClass = "hep-th",
    doi = "10.1007/JHEP05(2011)036",
    journal = "JHEP",
    volume = "05",
    pages = "036",
    year = "2011"
}

@article{Arias_2018,
   title={Entropy and modular Hamiltonian for a free chiral scalar in two intervals},
   volume={98},
   ISSN={2470-0029},
   url={http://dx.doi.org/10.1103/PhysRevD.98.125008},
   DOI={10.1103/physrevd.98.125008},
   number={12},
   journal={Physical Review D},
   publisher={American Physical Society (APS)},
   author={Arias, Raúl E. and Casini, Horacio and Huerta, Marina and Pontello, Diego},
   year={2018},
   month=Dec }

@article{flam,
  title = {R\'enyi Mutual Information in Quantum Field Theory},
  author = {Kudler-Flam, Jonah},
  journal = {Phys. Rev. Lett.},
  volume = {130},
  issue = {2},
  pages = {021603},
  numpages = {6},
  year = {2023},
  month = {Jan},
  publisher = {American Physical Society},
  doi = {10.1103/PhysRevLett.130.021603},
  url = {https://link.aps.org/doi/10.1103/PhysRevLett.130.021603}
}

@article{Cardy_2013,
   title={Some results on the mutual information of disjoint regions in higher dimensions},
   volume={46},
   ISSN={1751-8121},
   url={http://dx.doi.org/10.1088/1751-8113/46/28/285402},
   DOI={10.1088/1751-8113/46/28/285402},
   number={28},
   journal={Journal of Physics A: Mathematical and Theoretical},
   publisher={IOP Publishing},
   author={Cardy, John},
   year={2013},
   month=June, pages={285402} }

@article{Barnes:2010jp,
    author = "Barnes, Edwin and Vaman, Diana and Wu, Chaolun and Arnold, Peter",
    title = "{Real-time finite-temperature correlators from AdS/CFT}",
    eprint = "1004.1179",
    archivePrefix = "arXiv",
    primaryClass = "hep-th",
    doi = "10.1103/PhysRevD.82.025019",
    journal = "Phys. Rev. D",
    volume = "82",
    pages = "025019",
    year = "2010"
}

@article{Witten:1998qj,
    author = "Witten, Edward",
    title = "{Anti de Sitter space and holography}",
    eprint = "hep-th/9802150",
    archivePrefix = "arXiv",
    reportNumber = "IASSNS-HEP-98-15",
    doi = "10.4310/ATMP.1998.v2.n2.a2",
    journal = "Adv. Theor. Math. Phys.",
    volume = "2",
    pages = "253--291",
    year = "1998"
}

@article{Maldacena:1997re,
    author = "Maldacena, Juan Martin",
    title = "{The Large $N$ limit of superconformal field theories and supergravity}",
    eprint = "hep-th/9711200",
    archivePrefix = "arXiv",
    reportNumber = "HUTP-97-A097, HUTP-98-A097",
    doi = "10.4310/ATMP.1998.v2.n2.a1",
    journal = "Adv. Theor. Math. Phys.",
    volume = "2",
    pages = "231--252",
    year = "1998"
}

@article{Ryu:2006bv,
    author = "Ryu, Shinsei and Takayanagi, Tadashi",
    title = "{Holographic derivation of entanglement entropy from AdS/CFT}",
    eprint = "hep-th/0603001",
    archivePrefix = "arXiv",
    reportNumber = "NSF-KITP-06-11, NSF-KITP-06-11",
    doi = "10.1103/PhysRevLett.96.181602",
    journal = "Phys. Rev. Lett.",
    volume = "96",
    pages = "181602",
    year = "2006"
}

@article{Ryu:2006ef,
    author = "Ryu, Shinsei and Takayanagi, Tadashi",
    title = "{Aspects of Holographic Entanglement Entropy}",
    eprint = "hep-th/0605073",
    archivePrefix = "arXiv",
    reportNumber = "NSF-KITP-06-31, KUNS-2021, NSF-KITP-06-31, KUNS-2021",
    doi = "10.1088/1126-6708/2006/08/045",
    journal = "JHEP",
    volume = "08",
    pages = "045",
    year = "2006"
}

@article{Hubeny:2007xt,
    author = "Hubeny, Veronika E. and Rangamani, Mukund and Takayanagi, Tadashi",
    title = "{A Covariant holographic entanglement entropy proposal}",
    eprint = "0705.0016",
    archivePrefix = "arXiv",
    primaryClass = "hep-th",
    reportNumber = "DCPT-07-13, KUNS-2069",
    doi = "10.1088/1126-6708/2007/07/062",
    journal = "JHEP",
    volume = "07",
    pages = "062",
    year = "2007"
}

@article{Gubser:1998bc,
    author = "Gubser, S. S. and Klebanov, Igor R. and Polyakov, Alexander M.",
    title = "{Gauge theory correlators from noncritical string theory}",
    eprint = "hep-th/9802109",
    archivePrefix = "arXiv",
    reportNumber = "PUPT-1767",
    doi = "10.1016/S0370-2693(98)00377-3",
    journal = "Phys. Lett. B",
    volume = "428",
    pages = "105--114",
    year = "1998"
}

@article{witten1998antidesitterspacethermal,
    author = "Witten, Edward",
    editor = "Bergstrom, L. and Lindstrom, U.",
    title = "{Anti-de Sitter space, thermal phase transition, and confinement in gauge theories}",
    eprint = "hep-th/9803131",
    archivePrefix = "arXiv",
    reportNumber = "IASSNS-HEP-98-21",
    doi = "10.4310/ATMP.1998.v2.n3.a3",
    journal = "Adv. Theor. Math. Phys.",
    volume = "2",
    pages = "505--532",
    year = "1998"
}

@article{Furukawa_2009,
  title = {Mutual Information and Boson Radius in a $c=1$ Critical System in One Dimension},
  author = {Furukawa, Shunsuke and Pasquier, Vincent and Shiraishi, Jun'ichi},
  journal = {Phys. Rev. Lett.},
  volume = {102},
  issue = {17},
  pages = {170602},
  numpages = {4},
  year = {2009},
  month = {Apr},
  publisher = {American Physical Society},
  doi = {10.1103/PhysRevLett.102.170602},
  url = {https://link.aps.org/doi/10.1103/PhysRevLett.102.170602}
}

@article{Calabrese_2009,
   title={Entanglement entropy of two disjoint intervals in conformal field theory},
   volume={2009},
   ISSN={1742-5468},
   url={http://dx.doi.org/10.1088/1742-5468/2009/11/P11001},
   DOI={10.1088/1742-5468/2009/11/p11001},
   number={11},
   journal={Journal of Statistical Mechanics: Theory and Experiment},
   publisher={IOP Publishing},
   author={Calabrese, Pasquale and Cardy, John and Tonni, Erik},
   year={2009},
   month=Nov, pages={P11001} }

@article{Calabrese_2011,
   title={Entanglement entropy of two disjoint intervals in conformal field theory: {II}},
   volume={2011},
   ISSN={1742-5468},
   url={http://dx.doi.org/10.1088/1742-5468/2011/01/P01021},
   DOI={10.1088/1742-5468/2011/01/p01021},
   number={01},
   journal={Journal of Statistical Mechanics: Theory and Experiment},
   publisher={IOP Publishing},
   author={Calabrese, Pasquale and Cardy, John and Tonni, Erik},
   year={2011},
   month=Jan, pages={P01021} }

@article{De_Nobili_2015,
   title={Entanglement entropy and negativity of disjoint intervals in CFT: some numerical extrapolations},
   volume={2015},
   ISSN={1742-5468},
   url={http://dx.doi.org/10.1088/1742-5468/2015/06/P06021},
   DOI={10.1088/1742-5468/2015/06/p06021},
   number={6},
   journal={Journal of Statistical Mechanics: Theory and Experiment},
   publisher={IOP Publishing},
   author={De Nobili, Cristiano and Coser, Andrea and Tonni, Erik},
   year={2015},
   month=June, pages={P06021} }

@incollection{FisherHartwig1968,
    author    = {Fisher, Michael E. and Hartwig, Robert E.},
    title     = {Toeplitz Determinants: Some Applications, Theorems, and Conjectures},
    booktitle = {Advances in Chemical Physics},
    volume    = {15},
    pages     = {333--353},
    year      = {1968},
    doi       = {10.1002/9780470143605.ch18}
}

@book{Cover2006,
  author    = {Cover, Thomas M. and Thomas, Joy A.},
  title     = {Elements of Information Theory},
  edition   = {2nd},
  publisher = {Wiley-Interscience},
  year      = {2006},
  isbn      = {978-0471241959}
}

@article{Headrick:2010zt,
    author = "Headrick, Matthew",
    title = "{Entanglement Renyi entropies in holographic theories}",
    eprint = "1006.0047",
    archivePrefix = "arXiv",
    primaryClass = "hep-th",
    reportNumber = "BRX-TH-619, BRX-TH 619",
    doi = "10.1103/PhysRevD.82.126010",
    journal = "Phys. Rev. D",
    volume = "82",
    pages = "126010",
    year = "2010"
}

@article{Holzhey_1994,
   title={Geometric and renormalized entropy in conformal field theory},
   volume={424},
   ISSN={0550-3213},
   url={http://dx.doi.org/10.1016/0550-3213(94)90402-2},
   DOI={10.1016/0550-3213(94)90402-2},
   number={3},
   journal={Nuclear Physics B},
   publisher={Elsevier BV},
   author={Holzhey, Christoph and Larsen, Finn and Wilczek, Frank},
   year={1994},
   month=Aug, pages={443–467} }

@article{Klich:2026mlr,
    author = "Klich, Israel and Rajagopal, Ravisankar",
    title = "{An elementary approach to the Szegő theorems}",
    eprint = "2610.12724",
    archivePrefix = "arXiv",
    primaryClass = "math-ph",
    month = "10",
    year = "2026",
    journal = {arXiv preprint}
}

@book{simon2005orthogonal1,
  title     = {Orthogonal Polynomials on the Unit Circle, Part 1: Classical Theory},
  author    = {Simon, Barry},
  series    = {AMS Colloquium Publications},
  volume    = {54},
  year      = {2005},
  publisher = {American Mathematical Society},
  address   = {Providence, RI}
}

@article{Czech:2012bh,
    author = "Czech, Bartlomiej and Karczmarek, Joanna L. and Nogueira, Fernando and Van Raamsdonk, Mark",
    title = "{The Gravity Dual of a Density Matrix}",
    eprint = "1204.1330",
    archivePrefix = "arXiv",
    primaryClass = "hep-th",
    doi = "10.1088/0264-9381/29/15/155009",
    journal = "Class. Quant. Grav.",
    volume = "29",
    pages = "155009",
    year = "2012"
}
\end{document}